\documentclass{emulateapj}
\usepackage{natbib}
\usepackage{graphicx}
\usepackage{times}
\usepackage{url}
\usepackage{xspace} 
\usepackage{booktabs} 

\newcommand {\apgt} {\ {\raise-.5ex\hbox{$\buildrel>\over\sim$}}\ }
\newcommand {\aple} {\ {\raise-.5ex\hbox{$\buildrel<\over\sim$}}\ }

\newcommand {\chandra} {{\it Chandra}\xspace}
\newcommand {\xmm} {{\it XMM-Newton}\xspace}
\newcommand {\nustar} {{\it NuSTAR}\xspace}
\newcommand {\spitzer} {{\it Spitzer}\xspace}

\newcommand {\swiftbat} {{\it Swift}~BAT\xspace}

\newcommand {\wise} {{\it WISE}\xspace}

\newcommand {\sixum} {$6$~$\mu$m\xspace}
\newcommand {\twelveum} {$12$~$\mu$m\xspace}
\newcommand {\Lsixum} {$L_{\rm 6\mu m}$\xspace}
\newcommand {\fsixum} {$\hat{f}_{\rm 6\mu m}$\xspace}
\newcommand {\nuLnu} {${\rm \nu} L_{\rm \nu}$\xspace}
\newcommand {\feka} {Fe~K$\alpha$\xspace}

\newcommand {\brnu} {$\mathrm{BR}_{\mathrm{Nu}}$\xspace}

\newcommand {\nh} {$N_{\mathrm{H}}$\xspace}
\newcommand {\loiii} {$L_{\rm [O\ {\scriptscriptstyle III}]}$\xspace}
\newcommand {\lx} {$L_{\mathrm{X}}$\xspace}
\newcommand {\lxobs} {$L^{\rm obs}_{\mathrm{X}}$\xspace}
\newcommand {\lxint} {$L^{\rm int}_{\mathrm{X}}$\xspace}

\newcommand {\Lsoftobs} {$L^{\rm obs}_{\mathrm{2-10\ keV}}$\xspace}
\newcommand {\Lsoftint} {$L^{\rm in}_{\mathrm{2-10\ keV}}$\xspace}
\newcommand {\Lhard} {$L_{\mathrm{10-40\ keV}}$\xspace}
\newcommand {\Lhardobs} {$L^{\rm obs}_{\mathrm{10-40\ keV}}$\xspace}
\newcommand {\Lhardint} {$L^{\rm in}_{\mathrm{10-40\ keV}}$\xspace}

\newcommand{\oii}{\mbox{[\ion{O}{2}]}\xspace}
\newcommand{\oiii}{\mbox{[\ion{O}{3}]}\xspace}
\newcommand{\nev}{\mbox{[\ion{Ne}{5}]}\xspace}
\newcommand{\halpha}{H$\mathrm{\alpha}$\xspace}
\newcommand{\hbeta}{H$\mathrm{\beta}$\xspace}

\newcommand{\typeii}{type 2\xspace}

\newcommand{\mir}{mid-IR\xspace}

\newcommand {\ergpersec} {erg~s$^{-1}$\xspace}
\newcommand {\fluxunit} {erg~s$^{-1}$~cm$^{-2}$\xspace}
\newcommand {\nhunit} {cm$^{-2}$\xspace}
\newcommand {\degrees} {$^{\circ}$\xspace}

\newcommand {\xspec}{{\sc Xspec}\xspace}
\newcommand {\torus}{{\sc BNTorus}\xspace}
\newcommand {\mytorus}{{\sc MYTorus}\xspace}
\newcommand {\zwabspow}{$\mathtt{ZWABS\cdot POW}$\xspace}
\newcommand {\cabszwabspow}{$\mathtt{CABS\cdot ZWABS\cdot POW}$\xspace}
\newcommand {\plcabs}{$\mathtt{PLCABS}$\xspace}
\newcommand {\pexrav}{$\mathtt{PEXRAV}$\xspace}

\newcommand {\modelM}{$\mathtt{Model\ M}$\xspace}

\newcommand {\modelT}{$\mathtt{Model\ T}$\xspace}

\newcommand {\sdssa} {SDSS~J0011+0056\xspace}
\newcommand {\sdssb} {SDSS~J0056+0032\xspace}

\newcommand {\sdssd} {SDSS~J1034+6001\xspace}
\newcommand {\sdsse} {SDSS~J0840+3838\xspace}
\newcommand {\sdssf} {SDSS~J0913+4056\xspace}
\newcommand {\sdssg} {SDSS~J1218+4706\xspace}
\newcommand {\sdssh} {SDSS~J1713+5729\xspace}
\newcommand {\sdssi} {SDSS~J1243--0232\xspace}
\newcommand {\sdssj} {SDSS~J0758+3923\xspace}

\newcommand {\sdssdShort} {1034+6001\xspace}
\newcommand {\sdsseShort} {0840+3838\xspace}

\newcommand {\sdssgShort} {1218+4706\xspace}
\newcommand {\sdsshShort} {1713+5729\xspace}
\newcommand {\sdssiShort} {1243--0232\xspace}

\begin{document}

\title{\nustar Reveals Extreme Absorption in \lowercase{$z<0.5$} Type 2 Quasars}

\author{G.~B.~Lansbury\altaffilmark{1}, P.~Gandhi\altaffilmark{1,2},
  D.~M.~Alexander\altaffilmark{1}, R.~J.~Assef\altaffilmark{3}, J.~Aird\altaffilmark{4}, A.~Annuar\altaffilmark{1},
  D.~R.~Ballantyne\altaffilmark{5}, M.~Balokovi\'c\altaffilmark{6}, F.~E.~Bauer\altaffilmark{7,8,9},
  S.~E.~Boggs\altaffilmark{10}, W.~N.~Brandt\altaffilmark{11,12}, M.~Brightman\altaffilmark{6},
  F.~E.~Christensen\altaffilmark{13}, F.~Civano\altaffilmark{14,15,16}, A.~Comastri\altaffilmark{17}, W.~W.~Craig\altaffilmark{13,18}, 
  A.~Del~Moro\altaffilmark{1}, B.~W.~Grefenstette\altaffilmark{6},
  C.~J.~Hailey\altaffilmark{19}, F.~A.~Harrison\altaffilmark{6},
  R.~C.~Hickox\altaffilmark{16}, M.~Koss\altaffilmark{20},
  S.~M.~LaMassa\altaffilmark{14}, B.~Luo\altaffilmark{11,12},
  S.~Puccetti\altaffilmark{21,22}, D.~Stern\altaffilmark{23},
  E.~Treister\altaffilmark{24}, C.~Vignali\altaffilmark{17,25}, 
  L.~Zappacosta\altaffilmark{22}, W.~W.~Zhang\altaffilmark{26}}

\affil{$^1$ Centre for Extragalactic Astronomy, Department of Physics, University of Durham, South Road,
  Durham DH1 3LE, UK; g.b.lansbury@durham.ac.uk}
\affil{$^2$School of Physics and Astronomy, University of Southampton,
  Highfield, Southampton SO17 1BJ, UK}
\affil{$^3$N\'ucleo de Astronom\'ia de la Facultad de Ingenier\'ia,
  Universidad Diego Portales, Av. Ej\'ercito Libertador 441, Santiago,
  Chile}
\affil{$^{4}$Institute of Astronomy, University of Cambridge,
  Madingley Road, Cambridge, CB3 0HA, UK}
\affil{$^5$Center for Relativistic Astrophysics, School of Physics,
  Georgia Institute of Technology, Atlanta, GA 30332, USA}
\affil{$^{6}$Cahill Center for Astrophysics, 1216 East California
  Boulevard, California Institute of Technology, Pasadena, CA 91125,
  USA}
\affil{$^7$Instituto de Astrof\'{\i}sica, Facultad de F\'{i}sica,
  Pontificia Universidad Cat\'{o}lica de Chile, 306, Santiago 22,
  Chile}
\affil{$^{8}$Millennium Institute of Astrophysics, Vicu\~{n}a Mackenna 4860, 7820436 Macul, Santiago, Chile}
\affil{$^9$Space Science Institute, 4750 Walnut Street, Suite 205,
  Boulder, Colorado 80301, USA}
\affil{$^{10}$Space Sciences Laboratory, University of California,
  Berkeley, CA 94720, USA}
\affil{$^{11}$Department of Astronomy and Astrophysics, 525 Davey Lab,
  The Pennsylvania State University, University Park, PA 16802, USA}
\affil{$^{12}$Institute for Gravitation and the Cosmos, The Pennsylvania
  State University, University Park, PA 16802, USA}
\affil{$^{13}$DTU Space-National Space Institute, Technical University of
  Denmark, Elektrovej 327, DK-2800 Lyngby, Denmark}
\affil{$^{14}$Yale Center for Astronomy and Astrophysics, Physics
  Department, Yale University, PO Box 208120, New Haven, CT
  06520-8120, USA}
\affil{$^{15}$Smithsonian Astrophysical Observatory, 60 Garden Street,
  Cambridge, MA 02138, USA}
\affil{$^{16}$Department of Physics and Astronomy, Dartmouth College, 6127 Wilder Laboratory, Hanover, NH 03755, USA}
\affil{$^{17}$INAF Osservatorio Astronomico di Bologna, via Ranzani 1, 40127 Bologna, Italy}
\affil{$^{18}$Lawrence Livermore National Laboratory, Livermore, CA
  94550, USA}
\affil{$^{19}$Columbia Astrophysics Laboratory, 550 W 120th Street,
  Columbia University, NY 10027, USA}
\affil{$^{20}$Institute for Astronomy, Department of Physics, ETH Zurich, Wolfgang-Pauli-Strasse 27, CH-8093 Zurich, Switzerland}
\affil{$^{21}$ASDC-ASI, Via del Politecnico, 00133 Roma, Italy}
\affil{$^{22}$INAF Osservatorio Astronomico di Roma, via Frascati 33,
  00040 Monte Porzio Catone (RM), Italy}
\affil{$^{23}$Jet Propulsion Laboratory, California Institute of
  Technology, 4800 Oak Grove Drive, Mail Stop 169-221, Pasadena, CA
  91109, USA}
\affil{$^{24}$Universidad de Concepci\'{o}n, Departamento de
  Astronom\'{\i}a, Casilla 160-C, Concepci\'{o}n, Chile}
\affil{$^{25}$Dipartimento di Fisica e Astronomia, Universit\`a degli Studi di Bologna, 
Viale Berti Pichat 6/2, 40127 Bologna, Italy}
\affil{$^{26}$NASA Goddard Space Flight Center, Greenbelt, MD 20771,
  USA}

\shortauthors{Lansbury et al.}

\begin{abstract}
The intrinsic column density ($N_{\mathrm{H}}$) distribution of quasars is poorly known. At the high obscuration end of the quasar population and for redshifts $z<1$, the X-ray spectra can only be reliably characterized using broad-band measurements which extend to energies above $10$~keV. Using the hard X-ray observatory \nustar, along with archival \chandra and \xmm data, we study the broad-band X-ray spectra of nine optically selected (from the SDSS), candidate Compton-thick ($N_{\mathrm{H}}>1.5\times 10^{24}$~\nhunit) \typeii quasars (CTQSO2s); five new \nustar observations are reported herein, and four have been previously published. The candidate CTQSO2s lie at $z<0.5$, have observed \oiii luminosities in the range $8.4<\log (L_{\rm [O\ {\scriptscriptstyle III}]}/L_{\rm \odot})<9.6$, and show evidence for extreme, Compton-thick absorption when indirect absorption diagnostics are considered. Amongst the nine candidate CTQSO2s, five are detected by \nustar in the high energy ($8$--$24$~keV) band: two are weakly detected at the $\approx 3$$\sigma$ confidence level and three are strongly detected with sufficient counts for spectral modeling ($\gtrsim 90$ net source counts at $8$--$24$~keV). For these \nustar-detected sources {\it direct} (i.e., X-ray spectral) constraints on the intrinsic AGN properties are feasible, and we measure column densities $\approx2.5$--$1600$ times higher and intrinsic (unabsorbed) X-ray luminosities $\approx10$--$70$ times higher than pre-\nustar constraints from \chandra and \xmm. 
Assuming the \nustar-detected type 2 quasars are representative of other Compton-thick candidates, we make a correction to the \nh distribution for optically selected type 2 quasars as measured by \chandra and \xmm for $39$ objects. With this approach, we predict a  Compton-thick fraction of $f_{\rm CT}=36^{+14}_{-12}$~$\%$, although higher fractions (up to $76\%$) are possible if indirect absorption diagnostics are assumed to be reliable.
\end{abstract}


\section{Introduction}
\label{Introduction}

Much of the cosmic growth of supermassive black holes is
thought to occur
during a phase of luminous, heavily obscured accretion: an obscured
quasar phase \citep[e.g.,][]{Fabian99,Gilli07,Treister09}. However, our current census of
obscured quasars appears highly incomplete.
While unobscured quasars were first
discovered over 50~years ago \citep{Schmidt63,Hazard63}, it is
only in the last decade that (radio-quiet) obscured quasars have been discovered
in large numbers
\citep[e.g.,][]{Zakamska03,Hickox07,Reyes08,Stern12,Assef13,Donoso13}. Furthermore,
it is only very recently that the
most heavily obscured Compton-thick (with absorbing column
  densities of $N_{\rm H}> 1.5\times
10^{24}$~\nhunit; hereafter CT) quasars have begun to be
robustly identified at X-ray energies
\citep[e.g.,][]{Comastri11,Gilli11,Gandhi14,Lanzuisi15a}.

Identifying and characterizing heavily obscured quasars is
important for various reasons. 
Firstly, many less luminous AGNs in the local
Universe appear to be CT ($\sim 20$--$30\%$ of the total population; e.g., \citealt{Risaliti99,Burlon11}).
While observational constraints are challenging for distant
quasars, a significant population of luminous CT AGNs are expected
from models of the cosmic X-ray background
(CXB) spectrum \citep[e.g.,][]{Comastri95,Gilli07,Treister09,Draper10,Akylas12,Ueda14}. 
Secondly, while the orientation-based unified
model \citep[e.g.,][]{Antonucci93,Urry95} can account for the relative fractions of
unobscured, obscured and CT AGNs
observed in the local Universe, it is unclear whether a unified model
or some evolutionary scenario \citep[e.g.,][]{Sanders88,Hopkins08} is
more appropriate at higher luminosities and redshifts. 
Indeed, the observed dependence of AGN obscuration
  on luminosity suggests a departure from the unified model (e.g.,
  \citealt{Ueda03,Simpson05, 
Treister10,Iwasawa12,AssefSubmitted, 
Buchner15,Lacy15}). 
The above issues can
be addressed using X-ray studies which aim to measure the column
density ($N_{\rm H}$) distribution and CT fraction
of obscured quasars, important components of CXB models and important tools for
understanding AGN models \citep[e.g.,][]{Fabian09,Draper10}.

X-ray studies of 
heavily obscured quasars are extremely challenging. For instance,
to-date very few optically selected obscured quasars (i.e.,
  ``\typeii'' quasars or ``QSO2s'';
the definition of this term is provided in Section \ref{definitions}) have been unambiguously confirmed
as CT using broad-band X-ray measurements extending to high energies
($>10$~keV; e.g., \citealt{Gandhi14}). Including the high-energy data is crucial.  
Firstly, the number of counts is inherently low at $<10$~keV, due to
photoelectric absorption of the X-ray continuum, which
restricts the accuracy of X-ray spectral modeling and may lead
to an underestimate of the absorbing column density and intrinsic luminosity. Secondly, important
diagnostic features can be missed if
the observed X-ray energy window is narrow. Such
features include the photoelectric absorption cut-off (e.g., at
$\approx 10$~keV for a $z=0.2$ AGN absorbed by $N_{\rm H}
=10^{24}$~cm$^{-2}$ gas), and features of Compton
reflection/scattering from cold, dense gas. The latter become
prominent when CT levels of photoelectric absorption deeply suppress the primary
continuum, revealing strong Fe~K$\alpha$ fluorescent line emission at $6.4$~keV and
a Compton reflection ``hump'' at $>10$~keV
\citep[e.g.,][]{George91}, and may arise from an extended structure
such as the torus of the unified model \citep[e.g.,][]{Ghisellini94}. 

{\it NuSTAR} \citep{Harrison13}, launched in June 2012, has further opened
our window on the X-ray spectra of obscured AGNs,
with sensitivity up to $78.4$~keV. As the first orbiting observatory to focus high-energy ($>10$~keV)
X-rays, it provides a two orders of magnitude improvement in
sensitivity and over an order of magnitude improvement in angular
resolution relative to the previous-generation $>10$~keV observatories. 
Recent studies have demonstrated that, in the case of heavily obscured
quasars, the most accurate constraints on the absorbing column density
and intrinsic X-ray luminosity come from a combination of both {\it
  NuSTAR} and {\it XMM-Newton}/{\it Chandra} data, which provide the
broadest possible energy band pass for X-ray spectral modeling \citep[e.g.,][]{Luo13,Balokovic14,DelMoro14,Lansbury14,Gandhi14}. 

In this paper, we extend the work of \citeauthor{Lansbury14}
(\citeyear{Lansbury14}; hereafter L14) and \citeauthor{Gandhi14}
(\citeyear{Gandhi14}; hereafter G14), using \nustar
to study the high-energy emission of SDSS-selected QSO2s which are
candidates for being CT (i.e., candidate ``CTQSO2s'').
The targets were initially selected based on \oiii~$\lambda$5007 line
emission \citep{Zakamska03,Reyes08}, thought to be an unbiased indicator of
intrinsic AGN power (e.g., \citealt[][]{Heckman05,LaMassa10}; but see also \citealt{Hainline13}), and
subsequently identified as CT candidates within the detection
capabilities of \nustar using the low-energy X-ray data available
\citep[e.g.,][]{Jia13}.
L14 looked at an exploratory sample of three $z=0.41$--$0.49$
candidate CTQSO2s: one was weakly detected and shown to have a high column density of $N_{\rm
  H}\gtrsim 5\times 10^{23}$~\nhunit; the remaining two were undetected but shown to have suppressed X-ray
luminosities in the high-energy regime, suggestive of CT
absorption. G14 showed the lower redshift object \sdssd (also known as
Mrk~34; $z=0.05$)
to have a column density and intrinsic power an order of magnitude
greater than those measured with the pre-\nustar X-ray data,
unambiguously revealing the object to be a CTQSO2.

We present new results for a further five targets, bringing the
\nustar-observed SDSS-selected candidate CTQSO2 sample to a total size of nine
objects.
For the brightest two sources we model the broad-band X-ray spectra, for
one weakly detected 
source we characterize the spectrum using the X-ray band ratio, and for
all targets (including non-detections)
we use the X-ray:mid-IR ratio to infer the intrinsic AGN properties.
The paper is organised as follows: Section \ref{sample} details the sample selection; Section \ref{data} describes the X-ray and
multiwavelength data,
along with data reduction and analysis procedures; Section
\ref{Results} presents the results of X-ray spectral and
multiwavelength analyses; and Section \ref{discussion} discusses the
results for the full sample of nine \nustar-observed candidate CTQSO2s
in the context of the parent QSO2 population, including an estimation
of the \nh
distribution and CT fraction for $z<0.5$. The cosmology adopted
is ($\Omega_{M}$, $\Omega_{\Lambda}$, $h$) = (0.27, 0.73, 0.71).
Uncertainties and limits quoted throughout the paper correspond to the $90\%$
confidence levels (CL), unless otherwise stated.

\section{The QSO2 Sample}
\label{sample}

\subsection{Definitions}
\label{definitions}

Quasars are rapidly accreting black holes which emit large amounts of
  radiation, and have luminosities which typically place them above the knee of AGN luminosity
  function. Multiple thresholds exist in the literature for separating quasars from less
  luminous AGNs (e.g., ``Seyferts''). According to the classical
  threshold of \citet{Schmidt83}, quasars are those objects with 
  absolute $B$--band magnitudes of $M_{B}<-23$.
  Thus far we have used the term ``obscured'' rather loosely,
  since it has different implications depending on the wavelength regime
  in question. In the optical band, objects are identified as obscured
  if they show narrow line emission
  without broad (e.g., \halpha or \hbeta) components, a result of the
  central broad line region being hidden from the observer. These
  objects are classed as type 2s, or QSO2s if the luminosity is at quasar levels (in type 1s
  the broad line components are visible). At X-ray
  energies, objects are identified as obscured or ``absorbed'' if their
  X-ray continua show evidence for being absorbed by gas along the
  line-of-sight, with column densities of $N_{\rm
    H}>10^{22}$~\nhunit. The objects in this work
  originate from a sample of optically-identified QSO2s
  \citep{Zakamska03,Reyes08}. Several X-ray studies at $<10$~keV have now provided evidence that these
  optically-identified QSO2s are also absorbed at X-ray energies, with many objects showing
  indirect evidence for being absorbed by column densities in excess of $N_{\rm
    H}=1.5\times 10^{24}$~\nhunit (i.e., CT columns;
  \citealt{Vignali06,Vignali10,Jia13}). In this paper we look at the
  {\it direct} evidence for CT absorption in these
  optically-identified QSO2s, from X-ray 
  analyses which incorporate spectral information at $>10$~keV.

\subsection{Sample Selection}
\label{selection}

\begin{figure}
\centering
\includegraphics[width=0.47\textwidth]{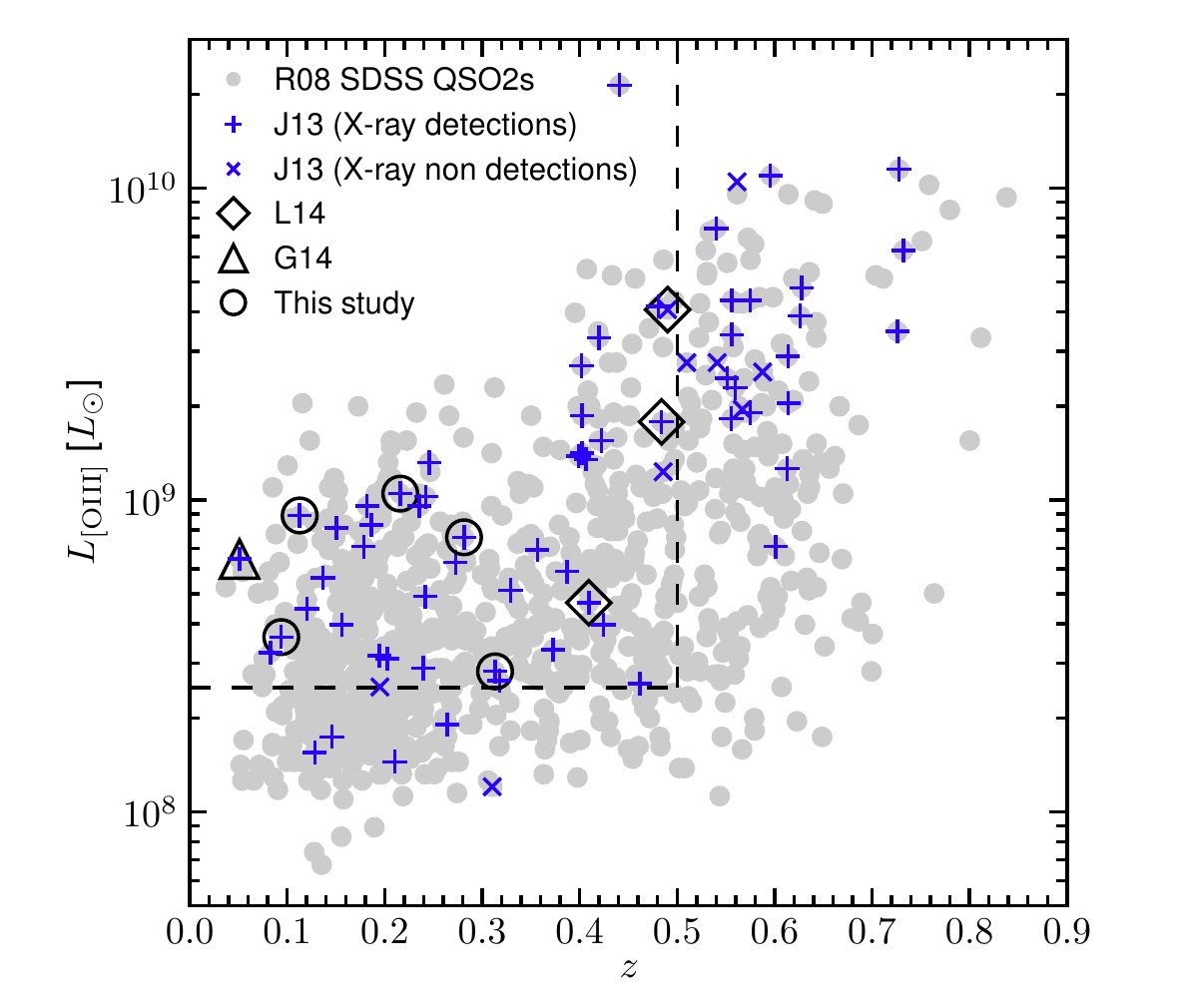}
\caption{Observed (i.e., extinction-uncorrected) \oiii~$\lambda$5007 line luminosity (\loiii) versus redshift
  ($z$). The R08 sample of SDSS-QSO2s is shown as
  grey dots. The J13 sample of \chandra- and \xmm-observed
  objects is indicated in blue, with `$+$'  and `$\times$' symbols
  indicating $<10$~keV detections and non detections, respectively
  (according to X-ray analyses in J13 and
  \citealt{Vignali06,Vignali10}).
  The dashed lines
  mark out the parameter space used in this work ($z<0.5$ and
\loiii$>2.5\times10^{8}L_{\odot}$), for
which the J13 sample is broadly representative
of the R08 sample. Our
  \nustar-observed subsample of candidate CTQSO2s is
  highlighted by black
  points, with circles marking the five recently observed objects
  presented in this study, diamonds marking the three $z\approx
  0.4$--$0.5$ objects presented in L14, and the triangle marking the low redshift ($z = 0.05$) object presented in
  G14.
}
\label{loiii_z}
\end{figure}

When selecting a sample of obscured quasars to observe at X-ray
energies, it is important to select based on an indicator of the
intrinsic AGN luminosity such that the sample is unbiased and as representative
of the general population as possible.
The \oiii$\lambda$5007 line, one of the strongest emission lines
readily visible in the optical, is a suitable choice since such
emission arises from gas on large ($\sim100$~pc) scales, minimizing
the effect of nuclear obscuration.
\citeauthor{Reyes08} (\citeyear{Reyes08}, hereafter R08; see also
\citealt{Zakamska03}) presented the largest sample of \oiii-selected
QSO2s, consisting of $887$ objects selected from the
SDSS. R08 defined
quasars as having observed (i.e., not
corrected for extinction) \oiii luminosities of
\loiii$>2\times10^{8}L_{\odot}$, and identified the quasars as type
2s (i.e., QSO2s) following the standard optical definition. For comparison,
the classical absolute magnitude cut of \citet[][$M_{B}<-23$]{Schmidt83}
corresponds approximately to \loiii$>3\times10^{8}L_{\odot}$ for type 1
sources \citep{Zakamska03}.
Subsequent {\it Chandra} and {\it XMM-Newton} studies
\citep[e.g.,][]{Ptak06,Vignali06,Vignali10,Jia13,LaMassa14} have investigated
the soft X-ray ($<10$~keV) properties of subsamples of the R08 sample,
with the largest subsample ($71$ objects) investigated by \citeauthor{Jia13} (\citeyear{Jia13}, hereafter
J13). 
Figure \ref{loiii_z} shows redshift versus \loiii for the R08 and J13
  samples.

For our study, we select from the J13 sample. In order
  to infer information about the overall optically selected QSO2
  population, we desire a parameter space for which the J13 sample is
  broadly representative of the R08 sample. As such we apply redshift and
  luminosity cuts of $z<0.5$ and \loiii$>2.5\times10^{8}L_{\odot}$,
  respectively (see Figure \ref{loiii_z}).
For these
$z$ and \loiii ranges: (1) the
$z$ and \loiii distributions of the J13 sample and
the R08 sample are consistent according to the
Kolmogorov-Smirnov (KS) test ($p=0.64$ and $0.09$ for $z$ and \loiii,
respectively); (2) the majority ($74\%$) of the J13 sample
are either serendipitous sources in the soft X-ray (\chandra
and \xmm) data or were targeted based on their \oiii properties, and
should therefore be relatively unbiased with respect to the X-ray
properties of the R08 sample.
We exclude \sdssf ($z=0.442$; \loiii$=2.1\times10^{10}L_{\odot}$), since
this infrared bright AGN is an extreme
outlier and has been targeted for \nustar separately (D.~Farrah et al., in preparation).
The above cuts leave $42$ QSO2s from J13, $39$ of which are
  detected at $<10$~keV (according to J13 and \citealt{Vignali06,Vignali10}).

From the J13 subsample above, we first targeted an initial three candidate CTQSO2s at
  $z\approx0.4$--$0.5$ (this subselection is described in L14). 
Since these three objects were weakly or not detected with \nustar,
for the succeeding targets described herein greater
consideration was given to the predicted
\nustar $8$--$24$~keV count rate.\footnote{The $8$--$24$~keV band is 
  the standard hard band defined for the \nustar extragalactic
  surveys \citep{Alexander13}.} 
The predictions were achieved by extrapolating from the
$<10$~keV data, assuming a variety of physically motivated torus models
which cover a range of column densities ($10^{23}<N_{\rm
  H}<10^{25}$~\nhunit).
To the remainder of the J13 subsample above, we applied a
  cut in observed X-ray:\oiii luminosity ratio of
$L^{\rm obs}_{\mathrm{2-10\ keV}}/$\loiii$<1$ (a conservative
threshold for targeting the most obscured candidates; see section 4.5 in J13), which leaves $12$ CT candidates.
From this selection, six objects were observed with \nustar, with
preference being given to the objects with high $8$--$24$~keV count
rate predictions. These include the one object presented in G14 and
the five presented in this paper, bringing the \nustar-observed SDSS-selected
candidate CTQSO2 sample to a total size of nine objects.

In this work we present results for the five recently observed
candidate CTQSO2s
\sdssj, 0840+3838, 1218+4706, 1243--0232 and 1713+5729.
For the other four previously-studied objects (\sdssa, 0056+0032,
1034+6001 and 1157+6003) the detailed reductions and data analyses are presented in
L14 and G14. Redshifts and \oiii luminosities for the five new objects
are listed in Table \ref{obs_table}. 
\begin{table*}[]
\centering
\caption{X-ray Observation Log}
\begin{tabular}{lcccccccccc} 
\hline\hline \noalign{\smallskip}
\multicolumn{3}{l}{} & \multicolumn{4}{c}{\nustar Observations} &
\multicolumn{4}{c}{Soft X-ray Observations} \\
\noalign{\smallskip}
\cmidrule(rl){4-7} \cmidrule(rl){8-11} 
\noalign{\smallskip}
\multicolumn{1}{c}{Object Name} & $z$ & \loiii & Observation ID & UT Date &
$t_{\rm on}$ & $t_{\rm eff}$ & Observatory & Observation ID & UT Date & $t$ \\
\multicolumn{1}{c}{(1)} & (2) & (3) & (4) & (5) & (6) & (7) & (8) &
(9) & (10) & (11) \\
\noalign{\smallskip} \hline \noalign{\smallskip}
SDSS J075820.98+392336.0 & 0.216 & 9.02 & 60001131002 & 2014:255 & 48.3 &
41.2 & \xmm & 0305990101 & 2006:108 & 9.1 \\ 

 & & & & & & & & 0406740101 & 2006:295 & 14.2 \\ 

 SDSS J084041.08+383819.8 & 0.313 & 8.45 & 60001132002 & 2014:121 &
 50.5 & 38.4 & \xmm & 0502060201 & 2007:289 & 19.0 \\ 

 SDSS J121839.40+470627.7 & 0.094 & 8.56 & 60001135002 & 2014:145 &
 41.8 & 34.0 & \xmm & 0203270201 & 2004:153 & 40.8 \\ 

 & & & & & & & & 0400560301 & 2006:321 & 43.2 \\ 

SDSS J124337.34--023200.2 & 0.281 & 8.88 & 60001136002 & 2014:211 &
55.5 & 46.0 & \chandra & 6805 & 2006:115 & 10.0 \\ 

 SDSS J171350.32+572954.9 & 0.113 & 8.95 & 60001137002 & 2014:120 &
 54.5 & 45.3 & \xmm & 0305750401 & 2005:174 & 4.4 \\ 
\noalign{\smallskip} \hline \noalign{\smallskip}
\end{tabular}
\begin{minipage}[l]{1\textwidth}
\footnotesize
NOTE. -- (1): Full SDSS object name. 
(2): Redshift. 
(3): Gaussian fit \oiii~$\lambda$5007 line luminosity [$\log(L_{\rm [O\ {\scriptscriptstyle III}]}/L_{\odot})$], as reported in R08.
(4) and (5): \nustar observation ID and start date (YYYY:DDD), respectively.
(6): Total on-source time (ks). 
(7): Effective on-axis exposure time (ks). This is the net value for the
$3$--$24$~keV band, and at the celestial coordinates of the target,
after data cleaning. We have accounted for vignetting; despite the sources
  being ``on-axis'', there is a small loss of exposure
  due to the natural dither of the observatory.
(8), (9) and (10): Soft X-ray observatory with available data,
corresponding observation ID(s) and start date(s) (YYYY:DDD), respectively. 
(11): Net on-axis,
flaring-corrected exposure time(s) (ks). For \xmm, the quoted
value corresponds to the EPIC detector used with the longest net exposure time.
\end{minipage}
\label{obs_table}
\end{table*}
The low-energy ($<10$~keV) X-ray
spectra have previously been characterized by J13,
who fit the existing \chandra and \xmm data with absorbed power law
models. 
For \sdssg, the column density constrained by J13 using this {\it direct}
(i.e., X-ray spectral) approach is high, but less than CT ($N_{\rm
  H}=8.0^{+5.6}_{-4.1}\times 10^{23}$~\nhunit).
In the other four cases, the directly constrained column densities are comparatively low ($N_{\rm H}<3\times
10^{22}$~\nhunit). 
This is in strong disagreement with the extremely low X-ray:\oiii
ratios, which imply
CT absorption. J13 recognised this, and thus used
indirect diagnostics to estimate the absorption levels. The low \nh measurements
from direct spectral fitting can be explained as
due to a combination of the limited energy ranges of \chandra and
\xmm, low source counts, and (especially in the case of \sdssh; see Section \ref{sdss1713_spectral} for further details) strong
contamination at lower energies from other processes such as star formation,
AGN photoionization, or scattered AGN emission.
In the Appendix we give individual
object information for the five
candidate CTQSO2s presented in this paper,
including relevant multiwavelength properties and indicators of heavy absorption. In addition, we comment on the single
  \nustar-detected candidate CTQSO2 from the exploratory study of L14 (\sdssa), for which a
  close inspection of the soft X-ray data reveals strong \feka emission.

\section{Data}
\label{data}

This section details the pointed \nustar observations and data
analysis procedures for the five newly observed SDSS-selected candidate
CTQSO2s (Section \ref{nudat}), which bring the \nustar-observed sample to
a total of nine such objects. We also detail the archival \chandra
and \xmm data (Section \ref{low_xray}), which facilitate a broad-band X-ray analysis when
combined with the \nustar data. 
In addition, near-UV to mid-IR data from large-area surveys
are presented in order to characterize the spectral energy distributions (SEDs) of
the objects and disentangle AGN and host galaxy emission in
the mid-IR (Section \ref{multiwav}).

\subsection{NuSTAR Data}
\label{nudat}

The \nustar observatory is
sensitive at $3$--$78.4$~keV (\citealt{Harrison13}). The combination of the instrumental
background and decrease in effective area with increasing energy means
that $3$--$\approx$$24$~keV is the most useful energy band for faint sources.
\nustar consists of two telescopes (A and B), identical in design, the
respective focal plane modules of which are referred to as FPMA and FPMB.
The point-spread function (PSF) has a tight ``core''  of ${\rm FWHM}=18\arcsec$ and a half-power diameter of $58\arcsec$.

Table \ref{obs_table} provides details, including dates and exposure
times, for the most recent five \nustar observations of SDSS-selected
candidate CTQSO2s. The data were processed as for the L14
sample, using the \nustar Data Analysis Software (NuSTARDAS)
version 1.3.0. For the detected sources, the
{\sc nuproducts} task was used to extract spectra and response files. Following other recent \nustar studies
(\citealt{Alexander13}; L14; \citealt{Luo14}), we perform photometry
in the $3$--$24$~keV, $3$--$8$~keV, and $8$--$24$~keV bands. The
photometry is performed for each FPM separately and also for combined
FPMA+FPMB data (referred to hereafter as ``FPMA+B''), to increase sensitivity. For
source detection, we use prior knowledge of the SDSS coordinates
and calculate no-source probabilities assuming binomial statistics ($P_{\rm B}$),
defining non detections as $P_{\rm B}>1\%$ (i.e., $\lesssim 2.6\sigma$).
For non detections we calculate upper limits on the net source counts
using the Bayesian approach outlined in \citet{Kraft91}.
For a detailed description of the source detection and aperture photometry
procedures, we refer the reader to L14. 

\begin{figure}
\centering
\includegraphics[width=0.47\textwidth]{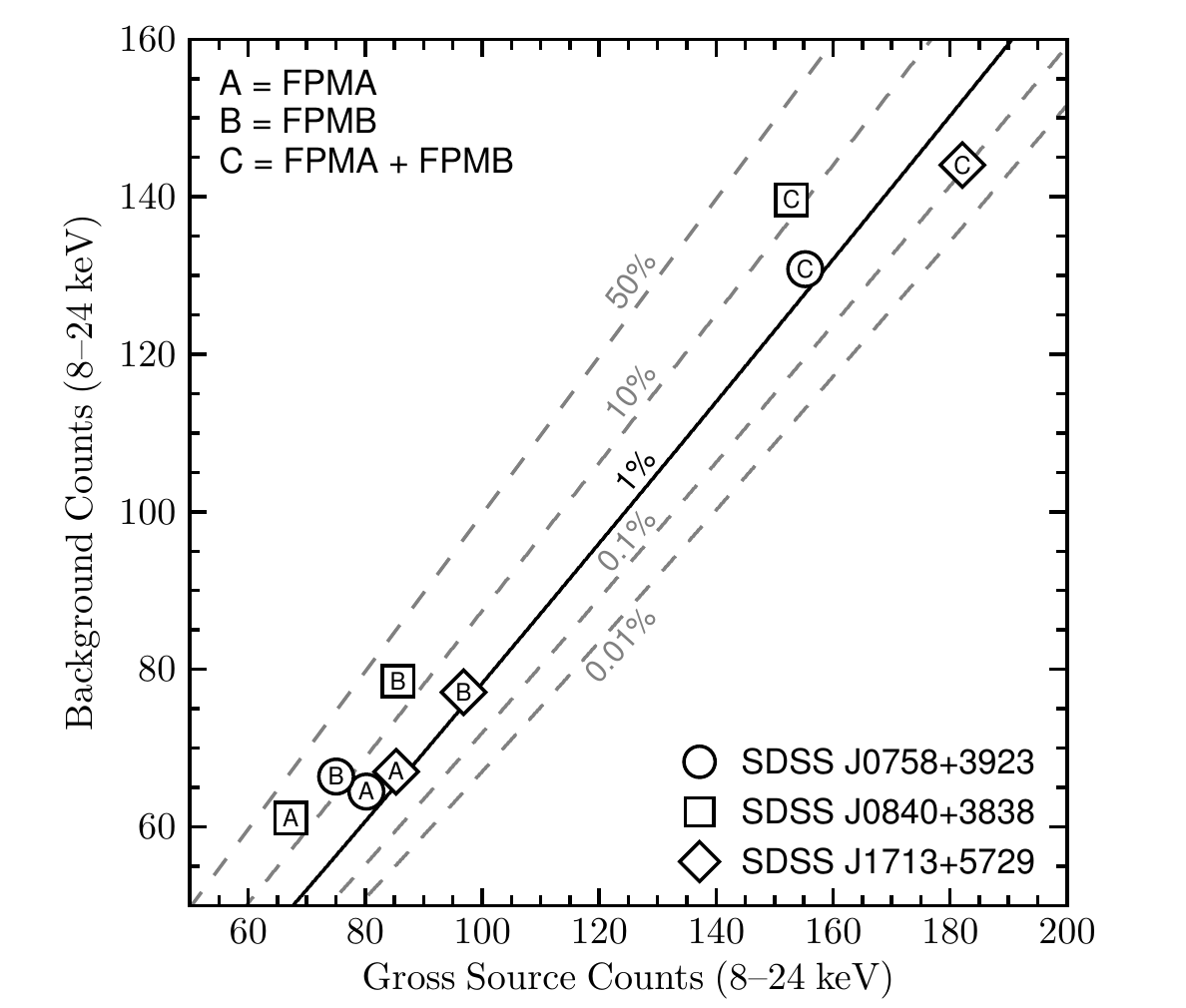}
\caption{\nustar photometry at $8$--$24$~keV for the faintest three sources, \sdssj, \sdsseShort and \sdsshShort
  (circles, squares and diamonds, respectively). Gross source counts and background counts
  (scaled to the source aperture) are shown. The dashed lines indicate tracks of constant Poisson no-source probability (a
 good approximation of $P_{\rm B}$, given the large background counts
 considered here; \citealt{Weisskopf07}). The solid black line shows
 our adopted detection threshold of $P_{\rm B}=1\%$. Only \sdssh is detected:
 while it is not detected in the individual FPMs, the increased
 sensitivity in FPMA+B (i.e., the combined FPMA+FPMB data) results in a
 significant detection, with $P_{\rm B}=0.22\%$. }
\label{false_probs}
\end{figure}

\begin{figure}
\centering
\includegraphics[width=0.47\textwidth]{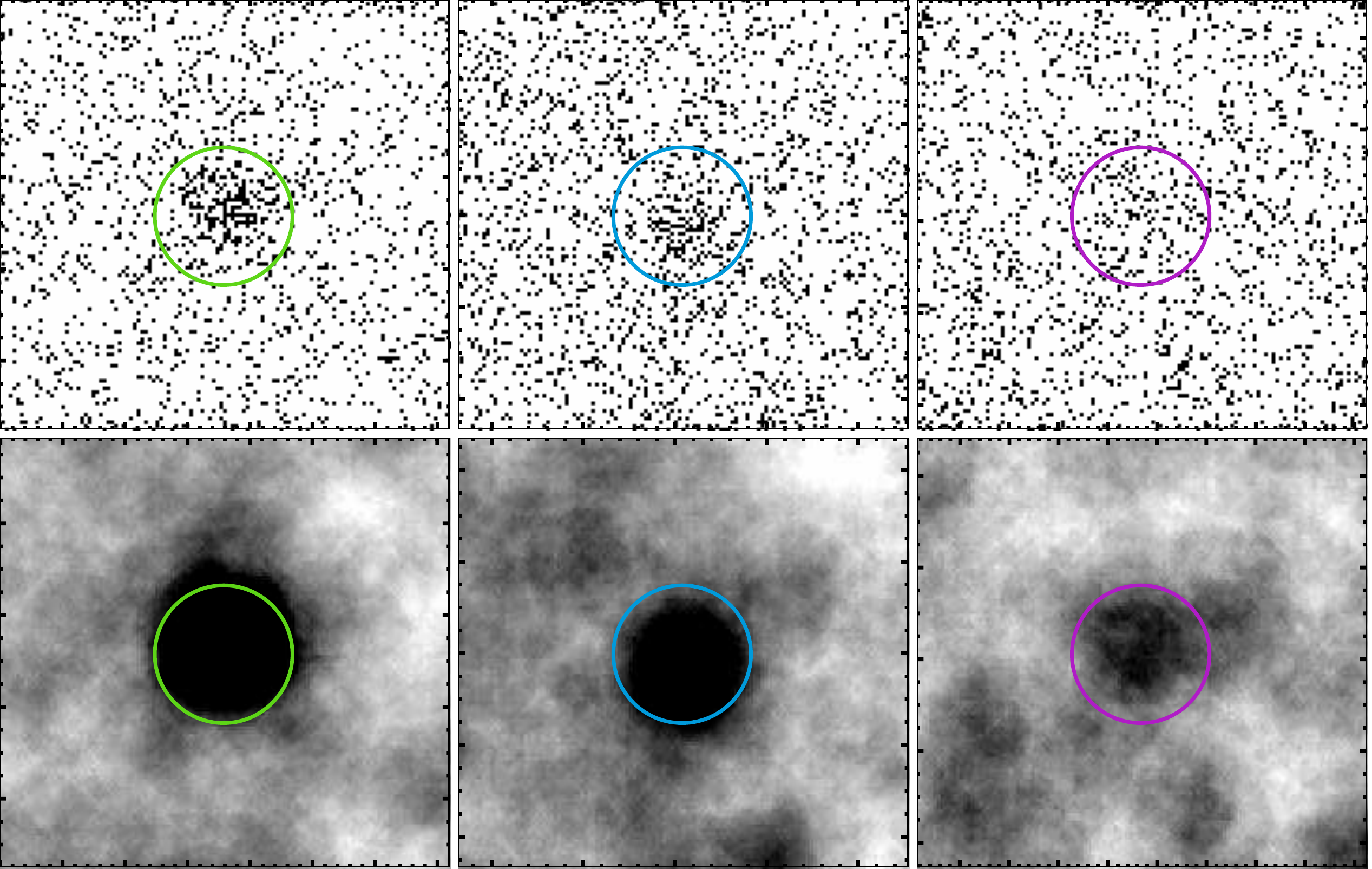}
\caption{\nustar $8$--$24$~keV images for the three objects detected
  in this energy band: \sdssg, \sdssiShort and \sdsshShort
  (left to right, respectively). Top row: unsmoothed photon
  images. Bottom row: images smoothed with a top hat function of radius
  14 pixels, corresponding to $34\farcs5$ (for aesthetic purposes
  only). The $45\arcsec$~radius source apertures are shown, centered on the SDSS positions.
The major
  tickmarks indicate 1 arcmin offsets in right ascension (R.A.; horizontal
  axis) and declination (Decl.; vertical axis).}
\label{nu_images}
\end{figure}

Table \ref{xray_photometry} summarizes the \nustar photometry.
\begin{table*}[]
\centering
\caption{X-ray Photometry: \nustar Counts}
\begin{tabular}{lccccccccc} \hline\hline \noalign{\smallskip}
\multicolumn{1}{c}{Object Name} & \multicolumn{3}{c}{Net Counts ($3$--$24$~keV)} & \multicolumn{3}{c}{Net Counts ($3$--$8$~keV)} & \multicolumn{3}{c}{Net Counts ($8$--$24$~keV)} \\
\noalign{\smallskip}
\cmidrule(rl){1-1} \cmidrule(rl){2-4} \cmidrule(rl){5-7} \cmidrule(rl){8-10} 
\noalign{\smallskip}
SDSS J & FPMA & FPMB & FPMA+B & FPMA & FPMB & FPMA+B  & FPMA & FPMB & FPMA+B  \\
\noalign{\smallskip} \hline \noalign{\smallskip}
 0758+3923 &  $30.4^{+17.9}_{-16.4}$ &  $<14.8$ &  $<43.8$ &  $<29.3$ &  $<7.2$ &  $<18.1$ &  $<30.4$ &  $<21.8$ &  $<45.0$ \\
 0840+3838 &  $<25.2$ &  $<17.1$ &  $<28.4$ &  $<14.6$ &  $<8.8$ &  $<13.4$ &  $<19.1$ &  $<21.5$ &  $<31.5$ \\
 1218+4706 &  $122.9^{+20.8}_{-19.3}$ &  $127.2^{+21.6}_{-20.2}$ &  $249.9^{+29.5}_{-28.0}$ &  $32.4^{+12.6}_{-11.1}$ &  $32.4^{+13.3}_{-11.8}$ &  $64.7^{+17.8}_{-16.4}$ &  $91.4^{+17.1}_{-15.6}$ &  $96.7^{+17.7}_{-16.2}$ &  $188.0^{+24.1}_{-22.6}$ \\
 1243-0232 &  $56.8^{+19.9}_{-18.4}$ &  $60.4^{+21.7}_{-20.2}$ &  $116.9^{+28.9}_{-27.5}$ &  $<32.4$ &  $<31.8$ &  $33.8^{+18.8}_{-17.3}$ &  $40.0^{+15.8}_{-14.3}$ &  $49.6^{+17.2}_{-15.7}$ &  $89.6^{+22.8}_{-21.3}$ \\
 1713+5729 &  $<43.1$ &  $<33.5$ &  $<67.4$ &  $<18.1$ &  $<13.3$ &
 $<21.5$ &  $<33.9$ &  $<36.3$ &  $38.1^{+19.6}_{-18.1}$ \\

\noalign{\smallskip} \hline \noalign{\smallskip}
\end{tabular}
\begin{minipage}[l]{1\textwidth}
\footnotesize
NOTE. -- \nustar net source counts for the candidate
CTQSO2s. FPMA and FPMB are the individual focal plane
modules belonging to the two telescopes which comprise \nustar. ``FPMA+B''
refers to the combined FPMA+FPMB data. 
\end{minipage}
\label{xray_photometry}
\end{table*}
Two of the quasars, \sdssg and \sdssiShort, are strongly detected; the net
source counts for FPMA+B in the $8$--$24$~keV band are $188$ and $90$,
respectively. Figure \ref{false_probs} shows the $8$--$24$~keV
no-source probabilities for the three fainter sources,
\sdssj, \sdsseShort and \sdsshShort. 
Poisson, rather than binomial, no-source
probabilities have been adopted for the purposes of the figure only,
to aid inter-object comparison; these provide a good approximation of
the binomial no-source probabilities ($P_{\rm B}$) since the
background counts are large \citep{Weisskopf07}. Although \sdssj is formally undetected at $8$--$24$~keV, it is only
just below the adopted detection threshold for this band and is weakly
detected in the
broader $3$--$24$~keV energy band, but for FPMA only ($P_{\rm
  B}=0.63\%$). \sdsse is a non detection. \sdssh is weakly detected
with FPMA+B for the $8$--$24$~keV band only ($P_{\rm B}=0.22\%$). 
In general, the detected sources have more net source counts in the $8$--$24$~keV band,
where the focusing soft X-ray observatories (e.g., \chandra and \xmm) have little to no sensitivity, than in the
$3$--$8$~keV band, where \nustar and the soft X-ray observatories overlap. This can occur for
heavily obscured AGNs, which have extremely flat X-ray
spectra and are therefore brighter at $\gtrsim 8$~keV. Indeed, the
single candidate CTQSO2 to be detected with \nustar
in L14, \sdssa, was only detected in the $8$--$24$~keV band.
\nustar FPMA+B $8$--$24$~keV image cutouts for the three new targets
detected in this energy band are shown in
Figure \ref{nu_images}.
None of these three sources are detected in the most sensitive
\swiftbat all-sky catalogs \citep[e.g.,][]{Baumgartner13}, and direct
examination of the 104 month \swiftbat maps shows no excess above
$2$$\sigma$ (for details of the maps and procedures, see
\citealt{Koss13}). Therefore, \nustar has provided the first real detections of these targets at
high energies ($>10$~keV).

For the \nustar-detected sources, it is important to rule out
confusion with and contamination from other nearby X-ray sources. 
Both of these are extremely unlikely: in the soft X-ray (\chandra and
\xmm) imaging of the \nustar-detected sources, the only neighbouring source
detected within $88\arcsec$ (i.e., the radial distance containing
an encircled-energy fraction of $\sim 85\%$ for the \nustar PSF) of the SDSS positions lies at an angular
separation of $51\arcsec$ from \sdssg (i.e., outside our adopted
source aperture radius) and is a factor of $\approx20$ fainter in the
\xmm energy band.

Table \ref{xray_lums} lists the aperture-corrected \nustar fluxes and rest-frame $10$--$40$~keV
luminosities (\Lhard; uncorrected for absorption). The
fluxes were obtained using photometry, assuming an effective
  photon index (i.e., for an unabsorbed power law model) of $\Gamma_{\rm eff}=0.3$ and using count rate to flux conversion factors
which account for the \nustar response and effective area. Often
$\Gamma_{\rm eff}=1.8$ (a typical value for the $3$--$24$~keV emission of AGNs; e.g.,
\citealt{Alexander13}) is assumed for such
  extrapolations, but the \nustar-detected candidate CTQSO2s have
  extremely flat observed spectral slopes at $3$--$24$~keV (see Section \ref{Results}),
  in agreement with $\Gamma_{\rm eff}=0.3$ in all cases.
For each object our measured \nustar flux is in agreement with the soft X-ray observatory (\chandra
or \xmm) measurement at $3$--$8$~keV, the energy band for which the
observatories overlap.
For the three faint or undetected sources (\sdssj, \sdsseShort and \sdsshShort), the \Lhard values were obtained by
extrapolating from the observed-frame $8$--$24$~keV fluxes assuming $\Gamma_{\rm eff}=0.3$. For the two
sources with good \nustar photon statistics (\sdssg and \sdssiShort) the
\Lhard values were calculated using the best-fitting spectral models (Section
\ref{xray_spectral}).

\begin{table*}[]
\centering
\caption{Multiwavelength Flux and Luminosity Measurements}
\begin{tabular}{lccccccccc} \hline\hline \noalign{\smallskip}
\multicolumn{1}{c}{Object} & \multicolumn{4}{c}{Observed-frame
  Flux ($10^{-13}$~\fluxunit)} & \multicolumn{3}{c}{Rest-frame
  Luminosity ($10^{42}$~\ergpersec)} & \multicolumn{1}{c}{$\hat{a}$} &
\multicolumn{1}{c}{\fsixum} \\
\noalign{\smallskip}
\cmidrule(rl){1-1} \cmidrule(rl){2-5} \cmidrule(rl){6-8} \cmidrule(rl){9-9} \cmidrule(rl){10-10} 
\noalign{\smallskip}
\multicolumn{1}{c}{} & \multicolumn{1}{c}{\chandra\ / {\it XMM}} &
\multicolumn{3}{c}{\nustar} &\multicolumn{1}{c}{\chandra\ / {\it XMM}}
& \multicolumn{1}{c}{\nustar} & \multicolumn{3}{c}{SED Modeling} \\
\noalign{\smallskip}
\cmidrule(rl){2-2} \cmidrule(rl){3-5}
\cmidrule(rl){6-6} \cmidrule(rl){7-7} \cmidrule(rl){8-10} 
\noalign{\smallskip}
\multicolumn{1}{c}{SDSS J} & $3$--$8$~keV &$3$--$8$~keV & $3$--$24$~keV & $8$--$24$~keV &
$2$--$10$~keV & $10$--$40$~keV & $6$~$\micron$ & $0.1$--$30$~$\micron$
& $6$~$\micron$ \\
\multicolumn{1}{c}{(1)} & (2) & (3) & (4) & (5) & (6) & (7) & (8) &
(9) & (10) \\
\noalign{\smallskip} \hline \noalign{\smallskip}

 0758+3923 &  $0.13^{+0.03}_{-0.02}$ &  $<0.12$ &  $<0.69$ &  $<0.93$ &  $2.33^{+0.40}_{-0.35}$ &  $<23.22$ &  $347\pm 19$ &  $0.88\pm 0.01$ &  $0.97\pm 0.00$ \\
 0840+3838 &  $<0.13$ &  $<0.09$ &  $<0.48$ &  $<0.69$ &  $<3.93$ &  $<35.23$ &  $130\pm 10$ &  $0.63\pm 0.03$ &  $0.73\pm 0.04$ \\
 1218+4706 &  $0.57^{+0.05}_{-0.47}$ &  $0.49^{+0.13}_{-0.12}$ &  $4.66^{+0.55}_{-0.52}$ &  $4.49^{+0.58}_{-0.54}$ &  $1.38^{+0.10}_{-1.13}$ &  $14.00^{+44.53}_{-1.17}$ &  $73\pm 3$ &  $0.91\pm 0.01$ &  $0.99\pm 0.01$ \\
 1243--0232 &  $0.15^{+0.08}_{-0.09}$ &  $0.19^{+0.11}_{-0.10}$ &  $1.65^{+0.41}_{-0.39}$ &  $1.62^{+0.41}_{-0.39}$ &  $5.74^{+0.69}_{-0.56}$ &  $54.60^{+5.22}_{-5.67}$ &  $25\pm 4$ &  $0.30\pm 0.04$ &  $0.52\pm 0.07$ \\
 1713+5729 &  $<0.30$ &  $<0.12$ &  $<0.95$ &  $0.69^{+0.35}_{-0.33}$ &  $<1.07$ &  $4.76^{+2.45}_{-2.26}$ &  $305\pm 21$ &  $0.92\pm 0.02$ &  $0.99^{+0.01}_{-0.03}$ \\

\noalign{\smallskip} \hline \noalign{\smallskip}
\end{tabular}
\begin{minipage}[l]{1\textwidth}
\footnotesize
NOTE. -- Columns ($2$) to ($7$): Hard X-ray (\nustar) and soft X-ray (\chandra or \xmm) fluxes
and luminosities. The rest-frame X-ray luminosities are observed values, i.e.\ uncorrected for 
absorption, and are in units of $10^{42}$~\ergpersec. 
The \nustar fluxes are from photometry in three observed-frame energy bands, assuming
$\Gamma_{\rm eff}=0.3$. The rest-frame $10$--$40$~keV luminosities are
determined from the best-fitting spectral models (Section
\ref{xray_spectral}) for \sdssg and
1243--0232, and by extrapolating from the observed-frame $8$--$24$~keV
band (assuming $\Gamma_{\rm eff}=0.3$) for \sdssj, 0840+3838 and 1713+5729.
The \chandra and \xmm fluxes and luminosities are determined from spectroscopy for \sdssj, 1218+4706 and
1243--0232, and from aperture photometry in the
observed-frame $3$--$8$~keV and rest-frame $2$--$10$~keV bands for \sdsse and 1713+5729
(assuming $\Gamma_{\rm eff}=0.3$). Columns ($8$) to ($10$): Best-fit
parameters from the near-UV to mid-IR SED modeling in Section \ref{multiwav}. The
errors shown correspond to standard deviations from a Monte Carlo
re-sampling of the photometric data. Column
($8$): rest-frame \sixum luminosity for the AGN only, \Lsixum
(\nuLnu), in units of $10^{42}$~\ergpersec. This value is intrinsic (i.e., corrected for
dust extinction). Column ($9$): The fractional contribution of the AGN to
the total integrated intrinsic luminosity between
$0.1$ and $30$~$\mu$m. Column ($10$): The fractional contribution of the AGN to
the observed (i.e., uncorrected for dust extinction) monochromatic rest-frame \sixum flux. 
\end{minipage}
\label{xray_lums}
\end{table*}

\subsection{Lower Energy X-ray Data}
\label{low_xray}

To incorporate lower energy ($<10$~keV; or ``soft'') X-ray data in our
study, we use archival \chandra and \xmm observations, limiting the analysis
to the $0.5$--$8$~keV and $0.5$--$10$~keV bands, respectively. Table
\ref{obs_table} provides details of the archival soft X-ray
observations, including dates and net exposure times.
For the sources with poor photon statistics, we perform photometry using
identical procedures to those for the \nustar photometry (see Section
\ref{nudat}). For the sources with good photon statistics, we model the X-ray
spectra with \xspec (see Section \ref{xray_spectral}). As mentioned in
Section \ref{nudat}, source confusion is extremely unlikely: there are no neighbouring sources
detected within $51\arcsec$ of the QSO2 positions. Measurements of the observed-frame $3$--$8$~keV fluxes
and rest-frame $2$--$10$~keV luminosities (uncorrected for 
absorption) are listed in Table \ref{xray_lums}.

For the source with \chandra coverage (\sdssi), we process the data
using {\sc chandra\_repro}.\footnote{http://cxc.harvard.edu/ciao/ahelp/chandra\_repro.html}
The source events are extracted from a circular $2\farcs5$ radius
aperture. The background events are extracted from a background
source-free annulus centered on the source coordinates, with an inner
radius of $8$\arcsec\ and an outer radius of $80$\arcsec. Since \sdssi
is on-axis, a large fraction ($\gtrsim 90\%$) of the source counts
lie within the source aperture. Given this, and the extremely low net
source counts measured ($9$), contamination of the background
region by source counts is negligible.

For the sources with \xmm coverage, we analyze data products from
the Pipeline Processing System (PPS) using the {\it Science Analysis
  Software} (SAS v.13.5.0).
To determine appropriate count rate thresholds for background flare subtraction, we
visually examine the light curves. In all cases the fraction of exposure time
removed is $\leq 30\%$, except in the case of obsID
0305750401 where the fraction is $49\%$. The exposure times in Table
\ref{obs_table} are flaring-corrected.
The source events are extracted from circular regions of $8$--$20$\arcsec\
radius (depending on source brightness and off-axis angle). The
background events are extracted from regions of area $70
\times 70$\arcsec\ to $140
\times 140$\arcsec, using either an annulus centered on the source position or
an offset region if it is necessary to avoid chip-gaps or nearby
sources. 
We combine the MOS1 and MOS2 data using the SAS task
{\sc epicspeccombine}, and simultaneously fit the PN and MOS data when
performing spectral analyses.

In the case of \sdssg, we use the two archival \xmm observations with the longest
exposures and most recent start dates (obsIDs 0203270201 and 0400560301). 
For obsID 0203270201, \sdssg lies close to the on-axis
position. In this instance we only use the MOS data, since
the source lies on a chip-gap for PN.
For obsid 0400560301, \sdssg lies far off-axis. In this case we only use the PN data, since the
source lies on a chip-edge in MOS1, and there are relatively low net counts with
MOS2 ($65$).


\subsection{Near-UV to Mid-IR SED Analysis}
\label{multiwav}

\begin{figure*}
\centering
\includegraphics[width=1.0\textwidth]{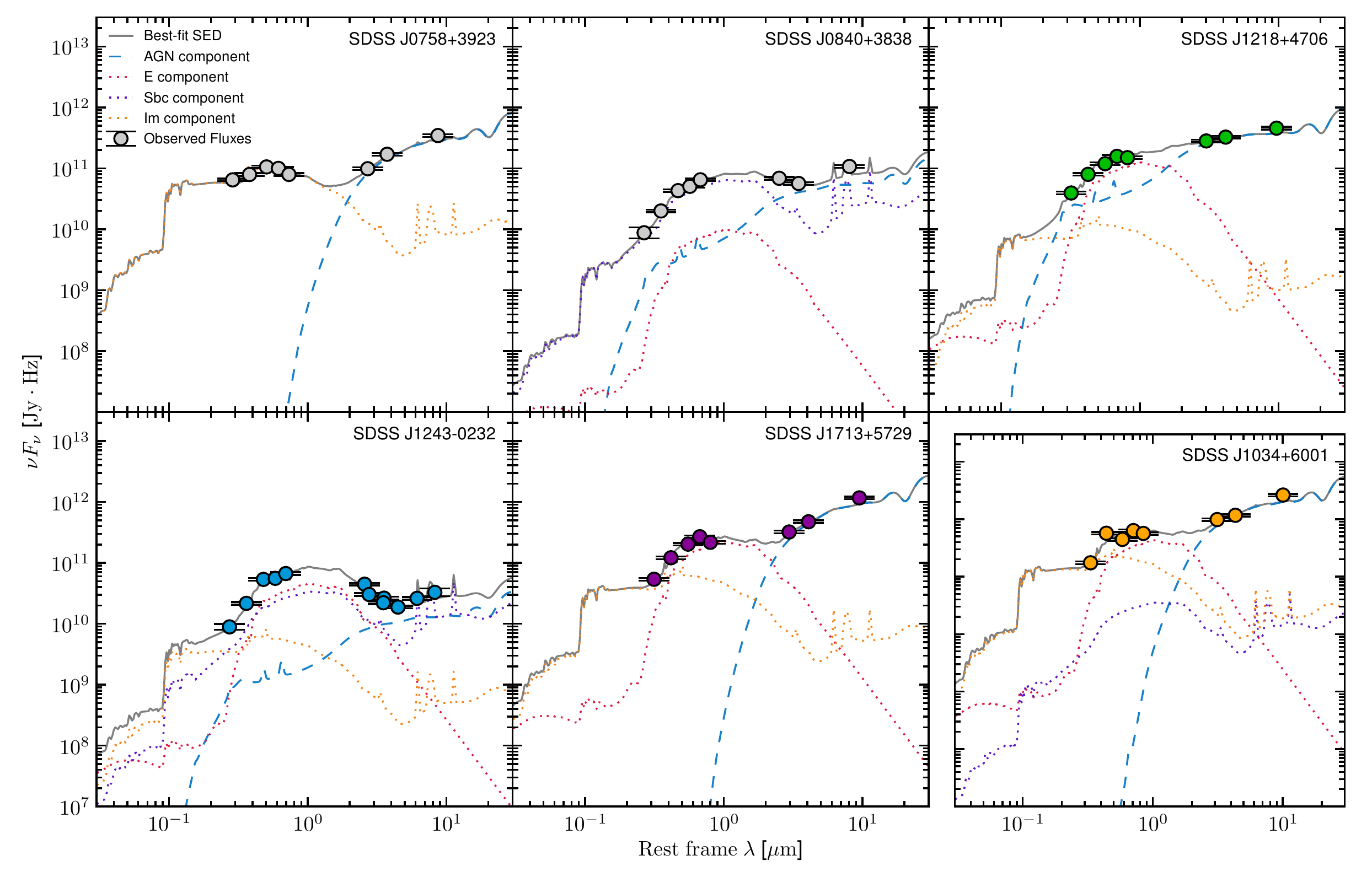}
\caption{Near-UV to mid-IR spectral energy distributions (SEDs) for
  the five candidate CTQSO2s presented in this work, and the one
  (\sdssd) presented in G14. AGN (blue dashed
  curve) and galaxy (dotted curves) templates were combined
  in the best-fit modeling of the photometric data (colored circles
 for the sources detected at $8$--$24$~keV with \nustar, and grey circles for the
 $8$--$24$~keV non-detections), following
  \citet{Assef08,Assef10,Assef13}. The three galaxy templates
  correspond to an old stellar population
(``elliptical'' or E; red), ongoing star formation (``spiral'' or
Sbc; purple),
and a starburst population (``irregular'' or Im; orange). The gray
curve shows the combined model
  solution. The systems are all AGN-dominated in the
  mid-IR waveband based on this analysis, except for \sdssi, which has comparable
  contributions from the AGN and the host galaxy; see Table \ref{xray_lums}.}
\label{sed}
\end{figure*}

Here we analyse near-UV to mid-IR ($0.3$--$30$~$\mu$m) spectral energy
distributions (SEDs) for the five candidate CTQSO2s
presented in this work, and the one presented in G14 (\sdssd), with the primary
aim of reliably measuring the AGN emission at mid-IR wavelengths. The photometric data (shown in Figure \ref{sed}) are collated from the SDSS \citep[Data
Release 7;][]{York00}, the {\it WISE} All-Sky source catalog
\citep{Wright10}, and the {\it Spitzer} \citep{Werner04} Enhanced Imaging Products
Source List (for \sdssi only). 
The SDSS fluxes are corrected for
Galactic extinction. The photometric data adopted are
  provided in the Appendix.
In order to provide a consistent SED analysis across the full sample of nine
\nustar-observed candidate CTQSO2s, we use the same SED decomposition
procedure as that applied in L14 to the initial three objects.
Following the methodology detailed in \citet{Assef08,Assef10,Assef13},
each SED is modeled as the best-fit, non-negative, linear combination of four empirical
templates \citep{Assef10}, including one AGN template and
three galaxy templates for: an old stellar population
(``elliptical'' or E), ongoing star formation (``spiral'' or Sbc),
and a starburst population (``irregular'' or Im). The internal dust
extinction of the AGN component is included as a free parameter in the
modeling. The
model solutions are shown in Figure \ref{sed}, and the following 
best-fitting parameters are listed in Table \ref{xray_lums}:
$\hat{a}$, the fractional contribution of the AGN to the total
intrinsic (i.e., corrected for the dust extinction of the AGN component) integrated $0.1$--$30$~$\mu$m
luminosity; \fsixum, the fractional contribution of the AGN to the
total observed (i.e., uncorrected for the dust extinction of the AGN component) monochromatic
rest-frame \sixum flux; and \Lsixum, the intrinsic AGN luminosity at
rest-frame \sixum (\nuLnu). The errors represent standard deviations from a Monte
  Carlo re-sampling of the photometric data over $1\,000$ iterations, and thus account for possible
  model degeneracies. In all cases the integrated light properties (i.e., the
  total galaxy and AGN contributions) are well constrained, which is
  required to accurately determine $\hat{a}$,
  \fsixum and \Lsixum. Since the primary goal of the SED modeling was
  to reliably measure these parameters, we do not make inferences about the host galaxy properties
  from the best-fit combination of
  host galaxy templates.
\sdssd, not shown in Table \ref{xray_lums} since the X-ray analysis is
presented in G14, has
\Lsixum$=(1.20\pm0.09)\times 10^{44}$~\ergpersec,
$\hat{a}=0.90\pm0.02$,
and \fsixum$=0.98^{+0.02}_{-0.03}$.

\begin{figure}
\centering
\includegraphics[width=0.47\textwidth]{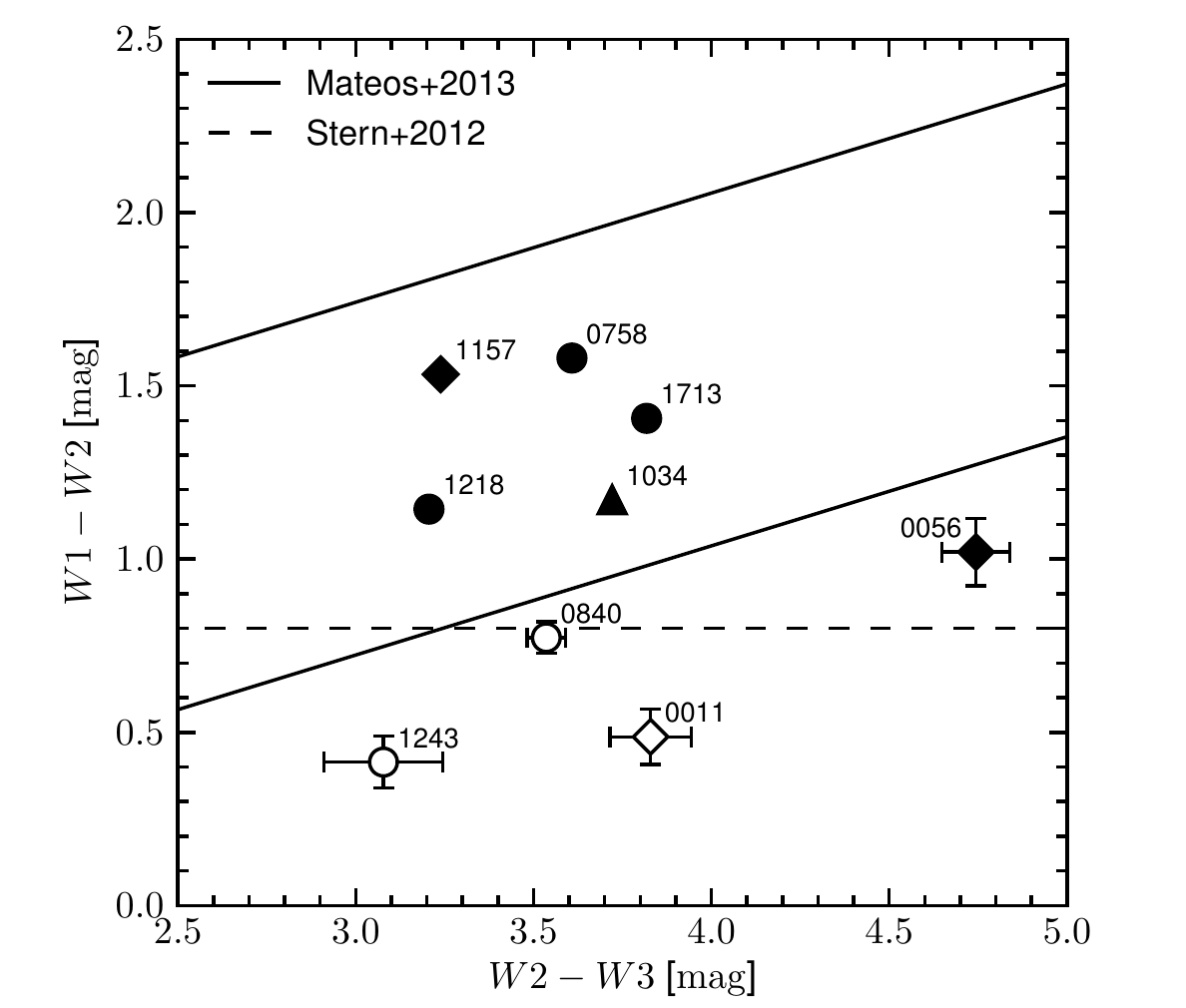}
\caption{\wise color--color diagram for the
  \nustar-observed SDSS-selected candidate CTQSO2s from this study (circles),
  G14 (triangle) and L14 (diamonds). We compare with the AGN
  color cut of \citeauthor[][]{Stern12} (\citeyear{Stern12}; $W{\rm
    1}$--$W{\rm 2}\geq0.8$) and the AGN `wedge' of
  \citet{Mateos13}. The filled and empty symbols mark sources which are strongly
  AGN-dominated ($\hat{a}\gtrsim 0.9$) and less AGN-dominated
  ($\hat{a}\lesssim 0.6$), respectively, at mid-IR wavelengths
  according to our SED modeling. 
For the five objects that lie within the
  AGN wedge, the error bars are smaller than the symbols.}
\label{wedge}
\end{figure}

The $\hat{a}$ constraints
demonstrate that the candidate
CTQSO2s in Figure \ref{sed} require an AGN component at a very
  high confidence level, and that in general the AGN contributes
strongly to the intrinsic emission across the broad
$0.1$--$30$~$\mu$m wavelength range (all but one object have $\hat{a}\gtrsim 0.6$). 
The high \fsixum values (all but one have \fsixum$\gtrsim 0.7$) indicate that the
observed monochromatic \sixum fluxes are AGN-dominated.
The presence of an AGN at mid-IR wavelengths may also be inferred
using \wise color diagnostics.
In Figure \ref{wedge} we show the six objects from Figure
\ref{sed}, and the three from L14, on the \wise $W{\rm
  1}$--$W{\rm 2}$ (i.e., [$3.4$~$\micron$]--[$4.6$~$\micron$]) versus
$W{\rm 2}$--$W{\rm 3}$ (i.e., [$4.6$~$\micron$]--[$12.0$~$\micron$])
plane. Generally, sources with larger $W{\rm
  1}$--$W{\rm 2}$ values have stronger AGN contributions. 
We compare with the AGN `wedge' of \citet{Mateos13} and the $W{\rm
  1}$--$W{\rm 2}$ color cut of \citet{Stern12}, which may be used to
identify AGN-dominated systems. Out of the total sample of nine candidate
CTQSO2s, five are AGN-dominated
according to both criteria, and one (\sdssb) falls below the
\citet{Mateos13} wedge but lies above the
\citet{Stern12} cut. This is in good agreement with the SED modeling
for these sources, where $\hat{a}\gtrsim 0.9$ in all cases. The
remaining three sources (\sdssa, \sdsseShort and \sdssiShort) fall below
both of the selection regions, although \sdsse is consistent with
satisfying the \citet{Stern12} AGN selection criterion given the errors. This
supports the SED modeling, from which it is concluded that these three sources are the least AGN
dominated ($\hat{a}\approx 0.3$--$0.6$, and $\hat{f}_{\rm 6\mu
  m}\approx 0.5$--$0.7$). The \wise colors of the objects agree
with the expectations; in general, the CTQSO2 population appears to
follow the \wise color distribution of the total QSO2 population, with a fraction of objects ($\sim
70\%$) lying within the AGN wedge \citep{Mateos13}. In the local
  Universe, $\sim40\%$ of the currently known bona fide CT AGNs lie within the wedge \citep{Gandhi15}.

In addition to the near-UV to mid-IR SED, one of the candidate CTQSO2s
presented in this work (\sdssh) has a detection at far-IR
wavelengths with {\it IRAS} which allows us to assess the extent to
which star formation could contribute to the soft X-ray emission
(Section \ref{sdss1713_spectral}).
\section{Results}
\label{Results}

To summarise the \nustar source detection for
the five SDSS-selected candidate CTQSO2s presented in this work: two are strongly
detected, one is weakly detected, and two are undetected by \nustar in
the high energy band ($8$--$24$~keV).
In Section \ref{xray_spectral} we present the results of X-ray
spectral fitting with \xspec for the three brightest objects. In
Section \ref{xray_br} we present the X-ray band ratios of all of the
\nustar-detected candidate CTQSO2s,
comparing to model predictions. For the weakly detected source \sdssh,
this is an appropriate method for
characterizing the broad-band X-ray spectrum. These two sections give
{\it direct} (i.e., X-ray spectral) constraints on absorbing column densities (\nh). In
Section \ref{indirect}, we present {\it indirect} constraints from a
multiwavelength diagnostic for the entire sample, including \nustar
non-detections.

\begin{figure}
\centering
\includegraphics[width=0.47\textwidth]{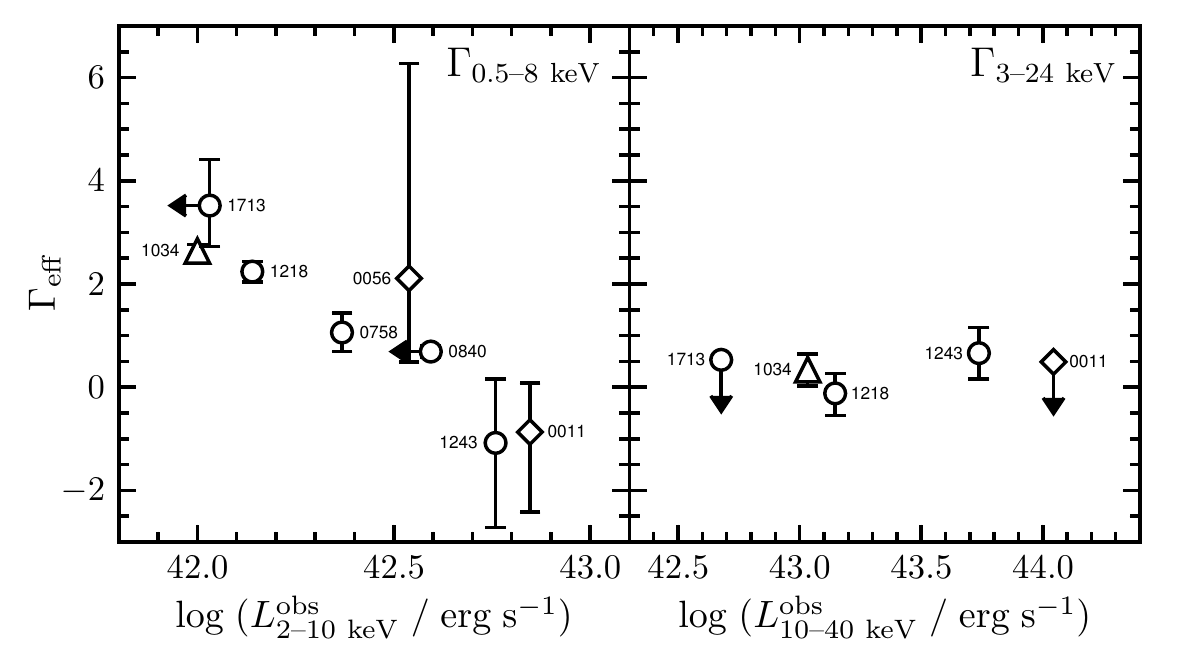}
\caption{Observed X-ray properties of the \nustar-observed candidate
  CTQSO2 sample. Left panel: properties measured at low
  energies with \chandra and \xmm. Right panel: properties
  measured at high energies with \nustar. Only detected sources are
  shown. Circles, diamonds and the triangle indicate the objects presented in
  this work, L14, and G14, respectively. The effective photon index ($\Gamma_{\rm eff}$), which
  provides a basic description of the overall X-ray spectral shape, was
  obtained by fitting an unabsorbed power law model to the data for
  each source. The rest-frame X-ray
  luminosities (\lxobs; bottom axis)
  are observed values, i.e. uncorrected for absorption. For
  the five objects presented in this paper, the luminosities
  correspond to those in Table \ref{xray_lums}. For the L14 and G14
  objects the luminosities have been calculated using the same
  methodology: spectral modeling where possible, or photometry
  following the procedure outlined in Section \ref{nudat}.
In the cases of the X-ray
  faint sources \sdssa and \sdssh,
  the $\Gamma_{\rm eff}$ for $3$--$24$~keV was estimated from the
  \nustar band ratio (\brnu; see Section \ref{xray_br}).}
\label{gam_lx}
\end{figure}

First we take a brief look at the overall X-ray spectral shapes for
the full sample of nine \nustar-observed candidate CTQSO2s. 
Figure \ref{gam_lx} shows the effective photon indices ($\Gamma_{\rm
  eff}$), measured through unabsorbed power law fits to the individual \chandra or \xmm ($0.5$--$8$~keV) and
\nustar ($3$--$24$~keV) spectra. 
The spectral shapes observed by \chandra and
\xmm vary significantly over an order of magnitude in
(non absorption corrected) rest-frame $2$--$10$~keV luminosity.
The increase in $\Gamma_{\rm eff}$~($0.5$--$8$~keV) toward lower
luminosities may reflect an increase in the relative contribution to the low-energy
spectra from processes unrelated to the direct AGN emission, such as thermal plasma
emission due to star formation or AGN-powered photoionization.
In contrast, the spectra seen by \nustar are consistent with
having the same effective photon index: excluding upper limits, the
mean is $\Gamma_{\rm eff}$~($3$--$24$~keV)~$\approx0.3$.

\subsection{X-ray Spectral Analysis: Best-fit Modeling}
\label{xray_spectral}

Here we use broad-band X-ray spectral modeling for the two
brightest \nustar-detected sources presented in this paper (\sdssg and \sdssi) to measure intrinsic properties: the intrinsic absorbing column
density (\nh), the intrinsic photon index ($\Gamma$), and the
intrinsic X-ray luminosity (\lx). 
Additionally, we investigate the low energy X-ray spectrum of \sdssh.
The X-ray spectral fitting is performed
using \xspec version 12.8.1j \citep{Arnaud96}.
In all cases we account for Galactic absorption using a
$\mathtt{PHABS}$ multiplicative component, with column densities fixed at values
from \citet{Kalberla05}.

\subsubsection{SDSS J121839.40+470627.7}
\label{sdss1218_spectral}

\begin{figure}
\centering
\begin{minipage}[l]{0.47\textwidth}
\includegraphics[width=\textwidth]{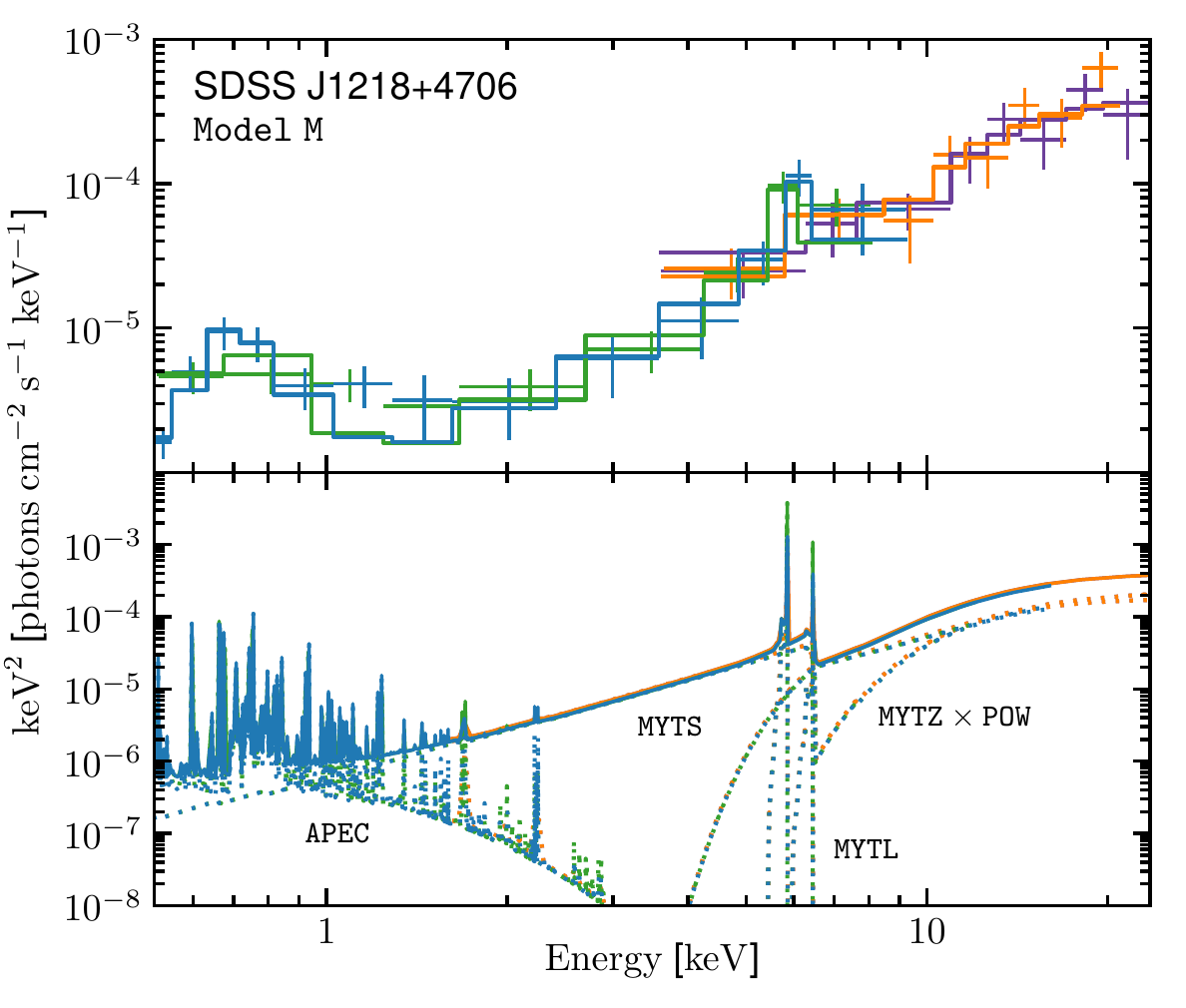}
\end{minipage}
\caption{Unfolded \nustar plus \xmm spectrum, in
  $EF_{E}$ units, for \sdssg. The data are shown in the upper panel, grouped to a minimum significance of $2\sigma$ per
bin for visual purposes. The \nustar FPMA and FPMB data are shown in purple and orange,
respectively. The MOS (obsID 0203270201) and PN (obsID 0400560301) data are shown in green and blue,
respectively. The best-fit \mytorus-based model (\modelM; described in
Section \ref{sdss1218_spectral}) is shown binned to match the data
(solid lines, upper panel)
  and in full detail (lower panel).}
\label{eeuf_1218}
\end{figure}

\sdssg has the strongest \nustar detection in the $8$--$24$~keV band, with net
source counts of $S_{\rm 8-24\ keV}=188$ for FPMA+B. 
The \nustar data are complemented by relatively high quality soft
X-ray data, with two long \xmm
exposures (obsIDs 0203270201 and 0400560301; see Table \ref{obs_table}).
Below we analyze the broad-band ($0.5$--$24$~keV) \nustar plus \xmm
dataset (shown in Figure \ref{eeuf_1218}). The modeling
  approach taken is similar to that adopted by G14 for \sdssd, the
  other brightest source in the \nustar-observed QSO2 sample, which
  has comparable photon statistics ($S_{\rm 8-24\ keV}=182$). We group the data by a minimum of $20$ counts per bin, and
use $\chi^{2}$ minimisation ($\mathtt{statistic\ chi}$ in \xspec) to constrain parameters. We note that using,
instead, $\mathtt{statistic\ cstat}$ (applying the
$W$~statistic approach; e.g., see Section \ref{sdss1243_spectral}) results
in essentially unchanged values for the key best-fit parameters ($\Gamma$ and \nh change by less than $0.1$ and $0.1\times
10^{24}$~\nhunit, respectively, for the models tested). The \xmm:\nustar
cross-normalization factor, when left as a free parameter, converges
to slightly different values depending on the model being tested, but
is always broadly
consistent (given the uncertainties) with the current best calibration measurements of \citet{Madsen15} of $\approx 0.93$. We therefore fix the cross-normalization factor to this
value throughout.

As shown in Figure \ref{gam_lx}, \sdssg has an extremely flat 
effective photon index over the \nustar band, $\Gamma_{\mathrm{3-24\ keV}} =
-0.15^{+0.40}_{-0.45}$. This is indicative of a
spectrum dominated by Compton reflection, as a result of the primary
continuum being heavily suppressed by CT levels of photoelectric
absorption \citep[e.g.,][]{George91}. 
Another important diagnostic feature of reflection is fluorescent
\feka line emission, which occurs at rest-frame $6.4$~keV and becomes
increasingly prominent as the level of
absorption increases \citep[e.g.,][]{Risaliti02}. An equivalent width threshold of $\mathrm{EW_{Fe\
    K\alpha}}>1$~keV is commonly used to identify CT AGNs; such
high values are difficult to explain for less than CT
columns \citep[e.g.,][]{Maiolino98a,Comastri04}, and
suggest a heavily reflection-dominated or pure
reflection spectrum, where little to none of the directly transmitted
AGN emission is visible.

For \sdssg, there is a clear excess of emission at observed frame
$\approx 6$~keV, which has previously been interpreted as 
  \feka line emission (J13; \citealt{LaMassa12a}).
To model this, 
we fit to the $>2$~keV \nustar plus \xmm dataset an unobscured power law
and Gaussian component, fixing the line energy at $E_{\mathrm{line}}=6.4$~keV and
the line width at $\sigma_{\mathrm{line}}=0.01$~keV. We measure an observed-frame equivalent
width of $\mathrm{EW_{Fe\ K\alpha}}=1.7^{+0.7}_{-0.6}$~keV using the \xmm spectra. 
This value is similar to but more tightly constrained than that published
by J13, since they only use one of the archival \xmm
observations, while we use two here.
The \feka line equivalent width is above the commonly adopted threshold
  for CT AGNs ($\mathrm{EW_{Fe\ K\alpha}}>1$~keV), with a comparable
  value to
that of the CT quasar \sdssd (Mrk~34; G14).
Freeing the Gaussian line energy
parameter, we obtain a best-fit value of $E_{\mathrm{line}}=6.40^{+0.24}_{-0.07}$~keV
(rest frame), which adds further confidence that the excess emission is due to \feka.

For the X-ray spectral modeling of \sdssg, we first conduct a simple test to assess
  the nature of the AGN continuum; we fit the $7$--$24$~keV \nustar data with two
  extreme models, one reflection-only spectrum and one transmission-only
  spectrum. Fitting the high energy data above $7$~keV allows a clean measurement
  of the AGN continuum {\it independent} of how the potentially complex lower energy
  emission is chosen to be modeled; low energy X-ray emitting processes other than the
  reflected or directly transmitted AGN continuum can dominate up to
  energies of $\approx 4$~keV
  \citep[e.g.,][]{Gandhi14,Gandhi15}, and fluorescent line
  emission (e.g., \feka) can also strongly contribute at energies up
  to $\approx 7$~keV. 
  For the reflection-only model we use \pexrav \citep{Magdziarz95}, with the reflection scaling factor set to $-1$ to
 produce a reflection-only spectrum (i.e., no directly transmitted
 component), and set all other parameters to the default values. 
This model provides a statistically acceptable fit to the \nustar
data ($\chi^{2}/n=11.3/12$; here, $n$ is the
number of degrees of freedom), and the intrinsic photon index is
constrained to be $\Gamma=1.35\pm0.46$.
For the transmission-only model we use \cabszwabspow (in
\xspec formalism).\footnote{The model \plcabs \citep{Yaqoob97} is generally a
  preferable transmission model to use (over \cabszwabspow) for 
  column densities of $N_{\rm H}>$~few~$\times
  10^{23}$~\nhunit. However, in our
 case \plcabs is not appropriate, since the energy range for which the model is valid depends on source column density
 ($E<14.4$~keV for $N_{\rm H} \leq 10^{24}$~\nhunit; $E<10$~keV for
 $N_{\rm H} \leq 5\times 10^{24}$~\nhunit; \citealt{Yaqoob97}), which
 means not utilising the high energy \nustar data.}
It is not possible to simultaneously constrain \nh and $\Gamma$ in
this case, so we fix the intrinsic photon index at $\Gamma=1.8$
\citep[a typical value for AGNs detected by \nustar at $3$--$24$~keV; e.g.,][]{Alexander13}. Again, there
is a statistically acceptable fit to the data
($\chi^{2}/n=10.5/12$), for a best-fit column density of $N_{\rm
  H} = (1.9^{+0.7}_{-0.5}) \times 10^{24}$~\nhunit. 

The above tests support the empirical evidence (from
  $\Gamma_{\rm eff}$ and $\mathrm{EW_{Fe\ K\alpha}}$) that extremely large, CT column densities
  are required to explain the X-ray spectrum of \sdssg. In the most
  extreme case, the source is consistent with being fully
  reflection-dominated (no directly transmitted component), which would
  imply $N_{\rm H}\gg 1.5 \times 10^{24}$~\nhunit. In the least extreme case, the
  source is consistent with lying close to the CT threshold ($N_{\rm
    H}\approx 1.5 \times 10^{24}$~\nhunit). However, the latter model
  assumes a transmission-only spectrum (no Compton reflection), which
  is unlikely given the large measured equivalent width of \feka. 
The reflection-only model tested (\pexrav) is also limited in that the
geometry (a slab of material) and infinite optical depth assumed are
not well motivated for obscured AGNs.
Ideally, in the CT regime, any absorbed continuum,
reflected continuum and fluorescent lines should be 
modeled in a self-consistent way, and assuming a well-motivated geometry. This is possible using the
physical models \mytorus \citep{Murphy09} and \torus \citep{Brightman11},
which were produced using Monte Carlo simulations of
X-ray radiative transfer through toroidal distributions of gas,
with the two models assuming different toroidal geometries.
We proceed to analyse the broad-band ($0.5$--$24$~keV) \xmm plus \nustar
spectrum of \sdssg using these two models.

\begin{table}[]
\centering
\caption{Best-fit Models for the \nustar +
  \xmm Spectrum of \sdssg}
\begin{tabular}{lcc} \hline\hline \noalign{\smallskip}
\multicolumn{1}{c}{} & \multicolumn{1}{c}{\modelM} &
\multicolumn{1}{c}{\modelT} \\
$\chi^{2}/n$ & $31.9/38$ & $33.0/39$ \\
$\Gamma$ & $2.4^{+0.2}_{-0.3}$ & $2.8^{+\mathrm{u}}_{-0.4}$ \\
$N_{\mathrm{H}}$ ($10^{24}$~$\mathrm{cm}^{-2}$) & $2.0^{+\mathrm{u}}_{-0.8}$ & $2.2^{+1.2}_{-0.6}$ \\
$\theta_{\rm tor}$ (\degrees) & [$60.0$] & [$60.0$] \\
$\theta_{\rm inc}$ (\degrees) & $63.7^{+8.5}_{-2.9}$ & [$87.0$] \\
$kT_{\mathtt{APEC}}$ (keV) & $0.42^{+0.20}_{-0.11}$ & $0.25^{+0.07}_{-0.05}$ \\
$L_{\rm 0.5-2keV}^{\mathtt{APEC}}$ ($10^{41}$~\ergpersec) & $1.38$ & $1.65$  \\
$L_{\rm 2-10keV}^{\rm obs}$ ($10^{44}$~\ergpersec) & $0.01$ & $0.01$  \\
$L_{\rm 10-40keV}^{\rm obs}$ ($10^{44}$~\ergpersec) & $0.14$ & $0.13$ \\
$L_{\rm 2-10keV}^{\rm int}$ ($10^{44}$~\ergpersec) & $0.85$  & $1.70$ \\
$L_{\rm 10-40keV}^{\rm int}$ ($10^{44}$~\ergpersec) & $0.46$ & $0.48$ \\
\noalign{\smallskip} \hline \noalign{\smallskip}
\end{tabular}
\begin{minipage}[l]{0.47\textwidth}
\footnotesize
NOTE. -- Best-fitting model parameters for the $0.5$--$24$~keV
spectrum of \sdssg. The individual models are detailed in Section
\ref{sdss1218_spectral}. The column
  densities (\nh) quoted are defined along the line-of-sight of the observer.
\end{minipage}
\label{model_table}
\end{table}
\begin{figure}
\centering
\includegraphics[width=0.47\textwidth]{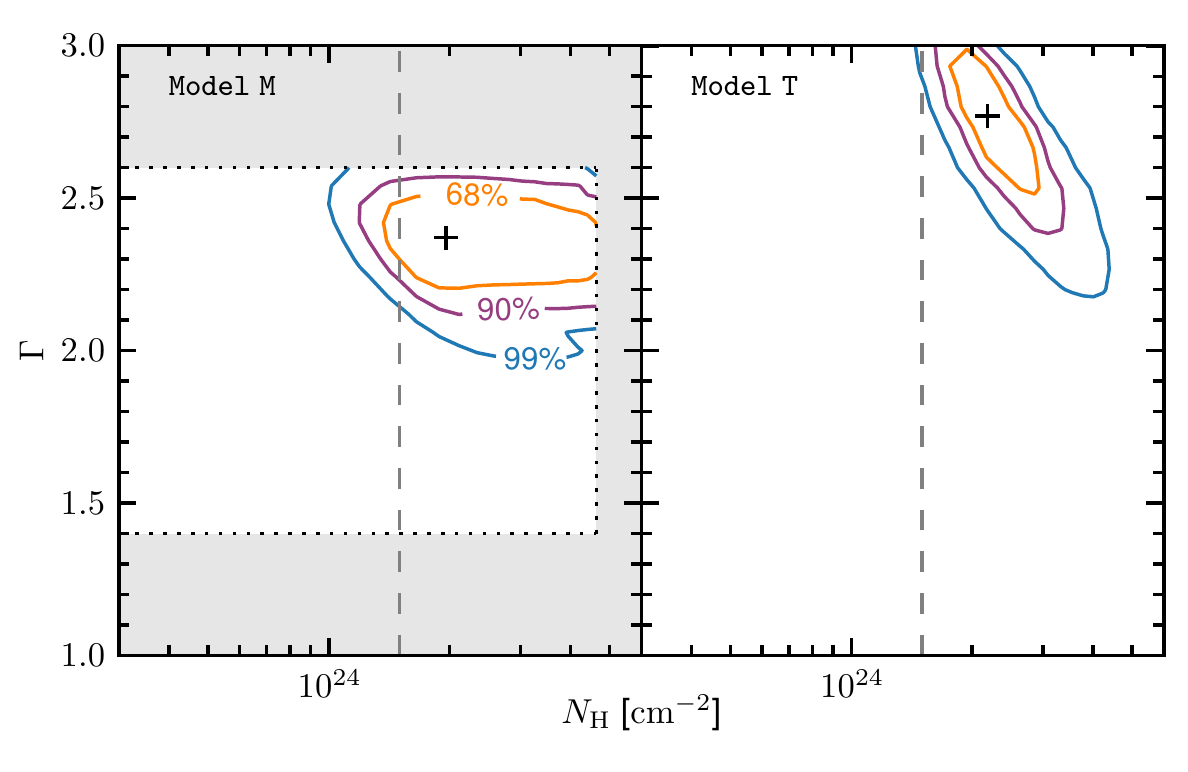}
\caption{Intrinsic photon index ($\Gamma$) versus (line-of-sight) column density ($N_{\mathrm{H}}$) confidence contours for
  \sdssg. 
The contours
  outline the $68\%$, $90\%$ and
  $99\%$ confidence regions, and the best-fit value is marked by a
  black cross. 
We show results for 
two models
  (\modelM and \modelT; left and right panels). The individual models are detailed in Section
  \ref{sdss1218_spectral}. The gray shaded region indicates the
  parameter ranges for
  which \modelM is not valid. The best-fit column densities are CT
  ($N_{\mathrm{H}}>1.5\times10^{24}$~\nhunit), and the $90\%$ CL
  lower $N_{\mathrm{H}}$ limits 
lie just below and just above the CT threshold (gray dashed line) for
  \modelM and \modelT, respectively.
}
\label{contours}
\end{figure}

Our \mytorus-based model (\modelM hereafter) has the following form:
\begin{flushright}
$\mathtt{Model\ M} = \mathtt{PHABS}\times (\mathtt{MYTZ}\times \mathtt{POW} + \mathtt{MYTS} + \mathtt{MYTL} + \mathtt{APEC})$.
\end{flushright}
Here, $\mathtt{MYTZ}$ reprocesses the zeroth-order transmitted
continuum ($\mathtt{POW}$) through photoelectric absorption and the
Compton scattering of X-ray photons out of the line-of-sight, $\mathtt{MYTS}$ is
the scattered/reflected continuum produced by scattering X-ray photons into
the line of sight, and $\mathtt{MYTL}$ is the fluorescent emission
line spectrum \citep{Murphy09}. We use \mytorus in the simplest form possible, tying the common parameters of
$\mathtt{MYTZ}$, $\mathtt{MYTS}$ and $\mathtt{MYTL}$ (\nh and
$\theta_{\rm inc}$) together. 
The intrinsic (unprocessed) photon indices and normalizations are tied to those of the
zeroth-order continuum ($\mathtt{POW}$). 
The torus opening angle ($\theta_{\rm tor}$) is fixed at $60$\degrees
in the current version of \mytorus.
$\mathtt{APEC}$ is a thermal plasma component \citep{Smith01} which we
use to parameterize the low energy excess, fixing the abundance
parameter at solar. This component is motivated by
the steep spectral slope at low energies ($\Gamma_{\mathrm{0.5-2\
    keV}}\approx 3.4$, measured using an unabsorbed power law model), which suggests contributions from processes
such as star formation or AGN photoionization, although we lack the spectral detail required
to distinguish between these processes.
The best-fit model has $\chi^{2}/n=32/38$ (see Table
\ref{model_table} for the model parameters and Figure \ref{eeuf_1218} for the
model spectrum).
Since $\Gamma$ and $N_{\mathrm{H}}$ are known to be
degenerate, we compute their uncertainties from $\chi^{2}$
contours in the $\Gamma$--$N_{\mathrm{H}}$ plane. Contours showing the
$68\%$, $90\%$ and $99\%$ confidence regions for this parameter space
are shown in Figure \ref{contours}. These were computed with
$\theta_{\rm inc}$ left free to vary.
Hereafter, the quoted uncertainties for \nh and $\Gamma$ are taken from
the $90\%$ CL contours.
The best-fit intrinsic photon index and line-of-sight column density are $\Gamma=2.4^{+0.2}_{-0.3}$ and $N_{\rm H}=(2.0^{+\mathrm{u}}_{-0.8})\times10^{24}$~cm$^{-2}$ [corresponding
to an equatorial column density of $N_{\rm H,eq}=(4.2^{+\mathrm{u}}_{-0.8})\times10^{24}$~cm$^{-2}$] for the best-fit inclination angle of
$\theta_{\rm inc}=63.7^{+8.5}_{-2.9}$$^{\circ}$. 
The modeling will not allow inclination angles of 
  $\theta_{\rm inc}<60$$^{\circ}$, since for these
  angles the observer has a direct, unobscured view of the central
  X-ray emitting source.
The upper error on \nh is not constrained,
which is in part due
to the limited \nh range of \mytorus ($N_{\rm H} = 10^{22}$--$10^{25}$~\nhunit). 
The best-fit model spectrum is reflection-dominated, with the
$\mathtt{MYTS}$ component dominating at $\approx 3$--$10$~keV, and the
$\mathtt{MYTZ}\cdot \mathtt{POW}$ and $\mathtt{MYTS}$ components contributing equally to the
normalization and spectral shape at $\gtrsim 10$~keV. 
To assess whether the \nustar plus \xmm spectrum is in agreement with being
fully reflection dominated, we test two modifications of \modelM where
the $\mathtt{MYTZ}\cdot \mathtt{POW}$ component is removed and the
inclination angle of the $\mathtt{MYTS}$ component is set to $0$\degrees and
$90$\degrees, corresponding to face-on and edge-on reflection.
Both models provide statistically acceptable fits to the spectrum ($\chi^{2}/n=29/35$
and $28/35$, respectively), with flat $\chi^{2}$ residuals, reasonable
best-fit intrinsic photon indices ($\Gamma=1.6^{+0.6}_{-\mathrm{u}}$ and $1.9^{+\mathrm{u}}_{-\mathrm{u}}$, respectively) and 
large column densities for the reflecting material [$N_{\rm H,reflector}=(3.1^{+\mathrm{u}}_{-1.6}) $ and $(1.5^{+1.0}_{-0.8}) \times10^{24}$~cm$^{-2}$, respectively].
The broad-band X-ray spectrum of \sdssg is therefore in agreement with being fully
reflection dominated. Since no transmission component is required in
these models, we may infer that the line-of-sight column density is
consistent with having a value of $N_{\rm H}\gg 1.5\times 10^{24}$~\nhunit.

Our \torus-based model (\modelT hereafter) has the following form:
\begin{flushright}
$\mathtt{Model\ T} = \mathtt{PHABS}\times (\mathtt{BNTORUS} + \mathtt{APEC})$.
\end{flushright}
In the \torus model, \nh is defined along the line of sight, and
is independent of $\theta_{\rm inc}$.
Initially, we fix the inclination at the maximum value of $\theta_{\rm
  inc}=87$\degrees, corresponding to an edge-on view of the torus.
Since the opening angle for \modelT is poorly constrained when left as
a free parameter ($\theta_{\rm tor}<72$\degrees), we fix it to $60$\degrees.
The best-fit model has $\chi^{2}/n=33/39$ (the model parameters are
listed in Table \ref{model_table}, and the $\Gamma$--\nh contours are
shown in Figure \ref{contours}). \nh is well constrained at the
$90\%$ CL, with a best-fit value of
$(2.2^{+1.2}_{-0.6})\times10^{24}$~\nhunit, and the intrinsic photon
index has a relatively high
value of $\Gamma=2.8^{+\mathrm{u}}_{-0.4}$. 
The upper error on $\Gamma$ is not constrained due to the parameter
limits of the \torus model. 
Fixing the intrinsic photon index at a more reasonable
  value of $\Gamma=2.3$, which is consistent with the $\chi^{2}$
  contours and is at the upper end of the range
  typically observed for unobscured AGNs
  \citep[e.g.,][]{Mateos10,Scott11}, results in a higher column
  density of $N_{\rm H}=(3.6^{+0.8}_{-0.7})\times10^{24}$~\nhunit and
  a reduced $\chi^{2}$ value close to unity ($\chi^{2}/n=39/40$).
 If the intrinsic photon index is fixed at $\Gamma=1.8$, an extremely high
 column density of $N_{\rm H}>5.1\times10^{24}$~\nhunit is required.
We note that the modeling (with $\Gamma$ left free) allows a large range of inclination angles ($\theta_{\rm
  inc}>63$\degrees), and re-modeling with $\theta_{\rm
  inc}$ fixed at a lower value of $65$\degrees results in a similarly good fit 
($\chi^{2}/n=38/39$) with no significant change in \nh but a flatter
photon index of $\Gamma=2.5^{+0.3}_{-0.4}$.
Furthermore, the statistical quality of the
fit and the best-fit parameters are relatively unchanged when $\theta_{\rm tor}$
is left as a free parameter.

To summarize, CT line-of-sight column densities 
are preferred for all of the models tested for \sdssg. The broad-band X-ray
spectrum shows evidence for
having a dominant contribution from Compton reflection, with the primary
continuum being heavily suppressed due to photoelectric
absorption. This is in
agreement with the expectations from the observation of strong
fluorescent \feka line emission
($\mathrm{EW_{Fe\ K\alpha}}\approx 1.7$~keV). The lowest limit
allowed by the modeling for the line-of-sight column
density is $N_{\rm H}>1.2\times
10^{24}$~\nhunit, and there is no constraint at the upper end. The
\nh, \lx and $\mathrm{EW_{Fe\ K\alpha}}$ constraints and data quality for \sdssg
($z=0.094$) are remarkably similar to those for the other low redshift
QSO2 strongly detected by \nustar, \sdssd ($z=0.051$; also known as
Mrk~34), which was identified by G14 as a bona fide CT AGN.
More complex models are possible (such as a clumpy torus; e.g., \citealt{Bauer14}), but
testing these is beyond the X-ray data quality.

\subsubsection{SDSS J124337.34--023200.2}
\label{sdss1243_spectral}
\sdssi is the third brightest \nustar detection in the SDSS-selected candidate
CTQSO2 sample, after \sdssg (Section \ref{sdss1218_spectral}) and \sdssd
(G14), but still has relatively low photon counts: $S_{\rm 8-24\
  keV}\approx 90$ and $S_{\rm 3-8\ keV}\approx 34$ with
\nustar, and $S_{\rm 0.5-8\ keV}\approx 9$ with \chandra. This emphasizes the
challenge involved in studying these inherently faint 
X-ray sources. Due to the low photon statistics, we 
use $\mathtt{statistic\ cstat}$ in \xspec, which is more appropriate than $\mathtt{statistic\ chi}$ in the case of
Poisson distributed data \citep{Nousek89}. 
In the case of unmodeled background spectra,
$\mathtt{cstat}$ applies the $W$~statistic 
\citep{Wachter79}.\footnote{See also
  http://heasarc.gsfc.nasa.gov/docs/xanadu/xspec/wstat.ps}
While the $W$~statistic is intended for unbinned data, bins containing
zero counts can lead to erroneous results,\footnote{See \url{https://heasarc.gsfc.nasa.gov/xanadu/xspec/manual/XSappendixStatistics.html}} so we group the \chandra and \nustar data by a minimum of $1$ count and $3$ counts per
bin, respectively \citep[e.g.,][]{Wik14}.
We fix the \chandra:\nustar cross-normalization factor at $1.0$,
consistent with the value obtained when the cross-normalization factor
is left as a free parameter in
the modeling.
\begin{figure}
\centering
\begin{minipage}[l]{0.47\textwidth}
\includegraphics[width=\textwidth]{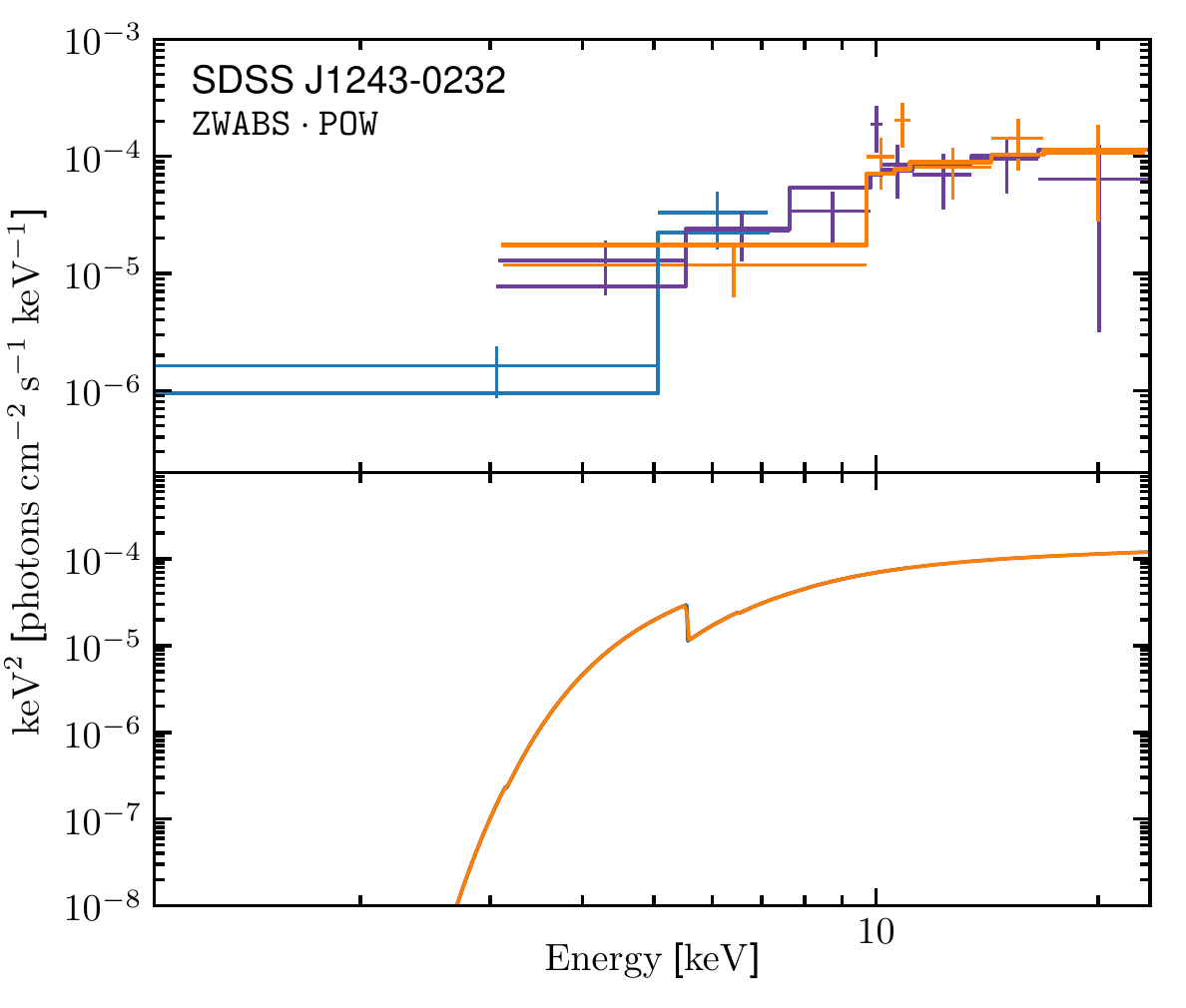}
\end{minipage}
\caption{Unfolded \nustar (purple and orange for FPMA and FPMB,
  respectively) plus \chandra (blue) X-ray spectrum for \sdssi. The
  best-fit absorbed power law (\zwabspow) model is shown. The
  panel layout, units and data binning follow that of Figure \ref{eeuf_1218}.}
\label{eeuf_1243}
\end{figure}

The \nustar spectrum of \sdssi
has a flat effective photon index of $\Gamma_{\mathrm{3-24\ keV}} =
0.66 \pm 0.50$, indicative of heavy absorption.
Fitting the broad-band ($0.5$--$24$~keV) \nustar plus \chandra
spectrum with a simple absorbed power law (\zwabspow) model, we obtain $N_{\rm
  H}\approx1.6\times10^{24}$~\nhunit and $\Gamma\approx 3$. This intrinsic photon index is discrepant with the expected range
for AGNs, and the parameter is poorly constrained. 
We therefore fix the parameter to $\Gamma=1.8$ (typical value in the
$3$--$24$~keV energy band for AGNs; e.g., \citealt{Alexander13}). 
The best-fitting model has $\chi^{2}=101$ and a C-statistic
  value of $C=123$, for $n=130$.  
The unfolded spectrum and best-fitting model are shown in Figure
\ref{eeuf_1243}.
The column density, $N_{\mathrm{H}} = (0.90^{+0.36}_{-0.33}) \times
10^{24}$~\nhunit, is close to CT. The intrinsic luminosities in the low and
high energy X-ray bands are 
$L^{\rm in}_{\mathrm{2-10\ keV}}=0.6\times 10^{44}$~\ergpersec and $L^{\rm in}_{\mathrm{10-40\ keV}}=0.7\times 10^{44}$~\ergpersec, respectively.
The higher quality \nustar data dominate the
fit, with similar results [$N_{\mathrm{H}} = (0.97^{+0.49}_{-0.38}) \times
10^{24}$~cm$^{-2}$] being obtained when the \chandra
data are excluded.
We note that $\mathtt{cstat}$ may also be used to model the unbinned,
gross (i.e., combined source plus background) spectrum, in which case the
Cash statistic ($C$~statistic; \citealt{Cash79}) is applied. 
Characterizing the background spectra using double power law models
($\mathtt{POW+POW}$ in \xspec), and including these as fixed
components in the spectral modeling of the \nustar data, this
$C$~statistic approach yields very similar results to the $W$~statistic approach, with $N_{\mathrm{H}} = (0.97^{+0.46}_{-0.37}) \times
10^{24}$~cm$^{-2}$.

Given the extremely flat effective photon index measured for this
  source, it is reasonable to
  test whether the spectrum is in agreement with a pure reflection
  continuum. As in Section \ref{sdss1218_spectral}, we use \pexrav
  with the reflection scaling factor set to $-1$ to produce a
  reflection-only spectrum. The model produces a similarly good fit to
  the data as for the absorbed power law model above, with $\chi^{2}=117$ and
  $C=120$, for $n=130$. We infer that the line-of-sight column density
  is consistent with being CT, with $N_{\rm H}\gg
  1.5 \times 10^{24}$~\nhunit. Unlike for the absorbed power law model, the intrinsic
  photon index is well constrained by the reflection-only model, with
  $\Gamma=1.7\pm0.3$. To summarize, the \nustar data unambiguously
  reveal heavy absorption in this QSO2, with a column density lower
  limit of $N_{\rm H} > 0.6\times 10^{24}$~\nhunit and no constraint at the high, CT
  absorption end. Higher quality X-ray data than those currently available,
especially at $<10$~keV, are required to reliably distinguish between
less than CT, and reflection-dominated CT models. For instance, the
current data are unable to provide informative constraints on 
\feka line emission (see the Appendix).

\subsubsection{SDSS J171350.32+572954.9}
\label{sdss1713_spectral}

For \sdssh there are too few \nustar counts for broad-band X-ray
spectral modeling (see Table \ref{xray_photometry}).
Here we investigate the low energy ($<10$~keV) spectrum observed with \xmm. The
object appears to have an extremely steep spectrum at low energies,
with PN (MOS) source counts of $<2$ ($<5$) at $2$--$10$~keV
and $12^{+6}_{-5}$ ($18^{+7}_{-5}$) at $0.5$--$2$~keV, implying a photon
index of $\Gamma=3.5_{-0.8}^{+1.0}$ in the $0.5$--$10$~keV energy
band; J13 measure a slightly flatter, but consistent (within the uncertainties), value
of  $\Gamma=2.5\pm 0.4$. 
The steep spectral slope is not typical of an AGN, 
and would be inconsistent with
the \nustar detection if produced as a result of direct AGN emission. 
To test whether the soft X-ray emission could be powered by star formation, we compare the $0.5$--$8$~keV luminosity,
$L_{\rm 0.5-8\ keV}=1.4\times 10^{42}$~\ergpersec, with 
the far-infrared (FIR) luminosity, $L_{\rm
  FIR}<4.0\times10^{44}$~\ergpersec, measured using {\it IRAS}
fluxes following \citet[][]{LonsdalePersson87}.
The relatively high soft X-ray:FIR luminosity ratio of $L_{\rm 0.5-8\ keV}/L_{\rm
  FIR}>0.0035$, which is a conservative lower limit due to the poorly
constrained {\it IRAS} $100\micron$ flux, rules out star
formation as the driver of the soft X-ray emission (e.g., see Figure 8 of \citealt{Alexander05}).
We deduce that the soft X-rays detected with \xmm are indirectly
powered by the AGN (e.g., via photoionization 
or scattered AGN emission), 
and \nustar may have provided the first identification of the directly
transmitted (or reflected) AGN continuum of this QSO2.

\subsection{X-ray Spectral Analysis: Band Ratios}
\label{xray_br}

\begin{figure}
\centering
\includegraphics[width=0.47\textwidth]{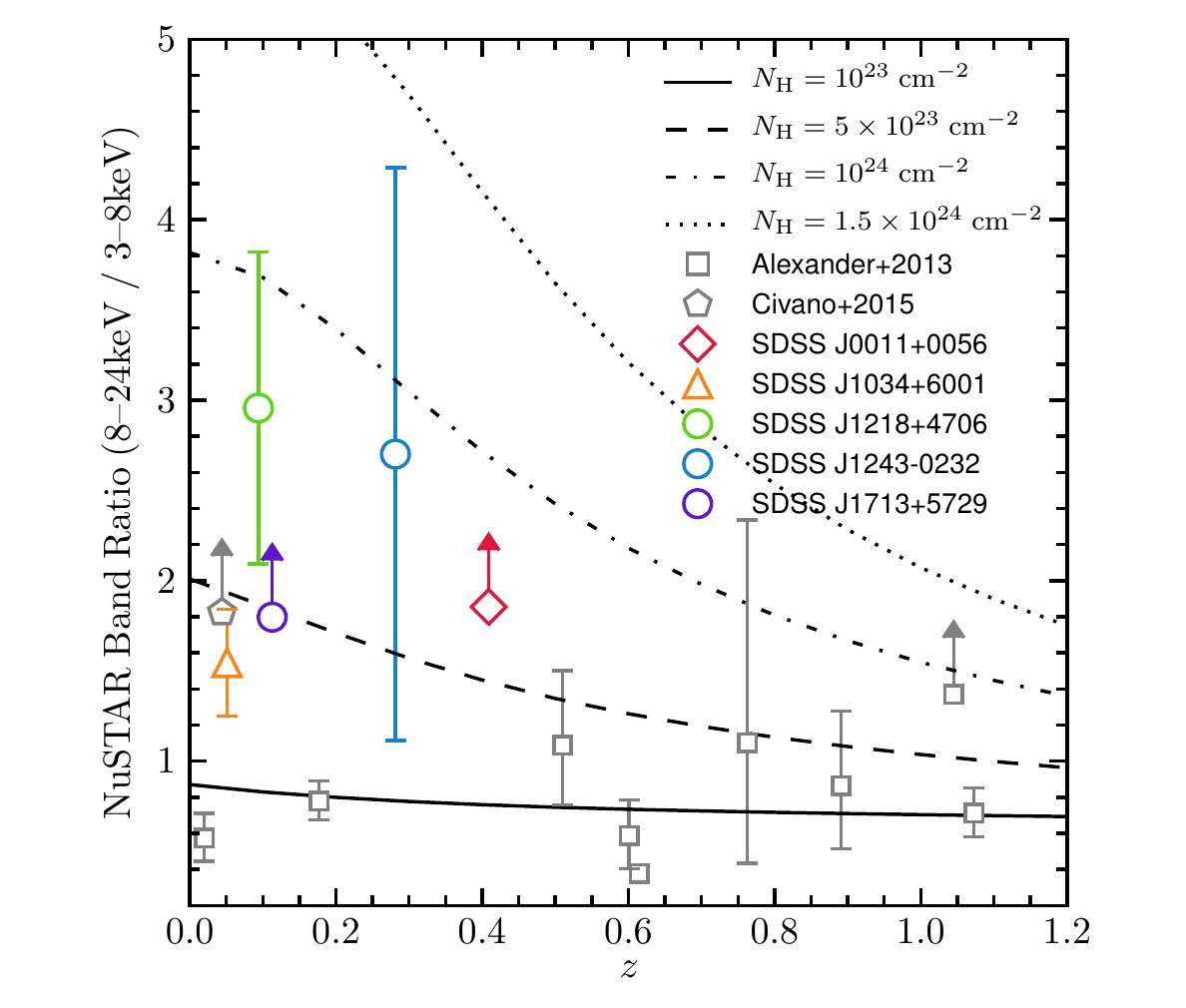}
\caption{\nustar band ratio (\brnu) versus $z$. The circles, the diamond and the
  triangle indicate the candidate CTQSO2s presented in this work, L14 and G14,
  respectively, which are detected at
  $8$--$24$~keV; $90\%$ CL error bars and limits are shown. For
  comparison, gray squares show
  the first $10$ AGNs detected in the \nustar serendipitous survey
  \citep{Alexander13}; $68\%$ CL error bars are shown. The gray
  pentagon marks a CT AGN identified with \nustar in the COSMOS field (F~Civano
  et al., submitted); a $90\%$ CL lower limit is shown. The tracks show
  model predictions for \brnu for four absorbing column densities in
  the range $N_{\rm H}=(0.1$--$1.5)\times10^{24}$~\nhunit.
The \brnu constraints for \sdssg, \sdssiShort and \sdsshShort are higher
than that of the confirmed CTQSO2 \sdssd, and suggest large absorbing columns.}
\label{br_z}
\end{figure}

X-ray band ratios provide a basic description of the X-ray
spectrum, and are useful when there are insufficient counts for detailed spectral
modeling. We define the \nustar band ratio (\brnu)
as the ratio of net source counts in the hard-band to those in
the soft-band, $S_{\rm 8-24\ keV}/S_{\rm 3-8\ keV}$. Figure \ref{br_z}
shows \brnu against redshift ($z$) for the five (of the total nine) \nustar-observed
candidate CTQSO2s which are detected at $8$--$24$~keV, including the
three presented in this paper (\sdssg, \sdssiShort and \sdsshShort)
and the two presented in L14 and G14 (\sdssa and \sdssdShort, respectively). 
The tracks show the expected evolution of
\brnu with $z$ for four different fixed column densities (\nh), computed
using a \mytorus model with an intrinsic photon index of $\Gamma=1.8$. 
We compare the measured \brnu values for the candidate CTQSO2s with these
tracks in order to infer \nh. 
We note that producing the tracks with,
instead, a simple \zwabspow
model results in higher \nh values for the same \brnu.
The \nustar-detected candidate CTQSO2s, in general, have high band ratios compared to
  AGNs detected in the \nustar extragalactic surveys (squares in Figure
  \ref{br_z}). In all cases the \brnu values suggest $N_{\rm H}>10^{23}$~\nhunit.

For \sdssh, a \nustar-detected object with too few counts for
broad-band spectral modeling of the direct AGN continuum (see Section
\ref{sdss1713_spectral}), the lower limit in \brnu suggests heavy
absorption with $N_{\rm H}\gtrsim5\times 10^{23}$~\nhunit. Our most
direct measurement for the intrinsic X-ray luminosity 
of this QSO2 comes from using this \nh constraint. Taking the 
observed $10$--$40$~keV luminosity constraint from Table \ref{xray_lums},
 and assuming that the X-ray spectrum is an absorbed power law with $\Gamma=1.8$,
the lower limits obtained are \Lsoftint$>4.6\times 10^{42}$~\ergpersec and
\Lhardint$>5.3\times 10^{42}$~\ergpersec.
As an alternative to the \brnu approach, \nh can be constrained using the \nustar/\xmm
band ratio (following L14). However, in this case the constraint ($N_{\rm H}\gtrsim2\times
10^{23}$~\nhunit) is less stringent than that
from \brnu, due to the comparatively poor quality of the available \xmm data.

The \nh estimates made from \brnu using Figure \ref{br_z} are relatively crude,
since the individual X-ray spectra may
have additional spectral complexities (e.g., line emission around $\approx 6.4$~keV, a scattered power law,
or a complex absorber geometry) not incorporated in our model
predictions. To illustrate this, for the two sources with
comparatively high quality \nustar spectra (\sdssd and \sdssgShort), the
less than CT column densities inferred from the \brnu analysis
($N_{\rm H} \lesssim 5 \times 10^{23}$~\nhunit and
$\lesssim 10^{24}$~\nhunit, respectively) are an
underestimate of the column densities determined from X-ray spectral
fitting ($N_{\rm H} \gtrsim 1.5 \times 10^{24}$~\nhunit; see G14 and
Section \ref{sdss1218_spectral} of this paper, respectively).
Similarly, using the \nustar results for three CT reflection-dominated
Seyfert~2s, \citet{Balokovic14} demonstrate that the above \brnu
approach underestimates \nh for reflection-dominated AGNs.
Nevertheless, \brnu provides first-order \nh constraints for weakly detected
sources.

\subsection{Indirect Constraints on X-ray Absorption}
\label{indirect}

\begin{figure*}
\centering
\includegraphics[width=1.0\textwidth]{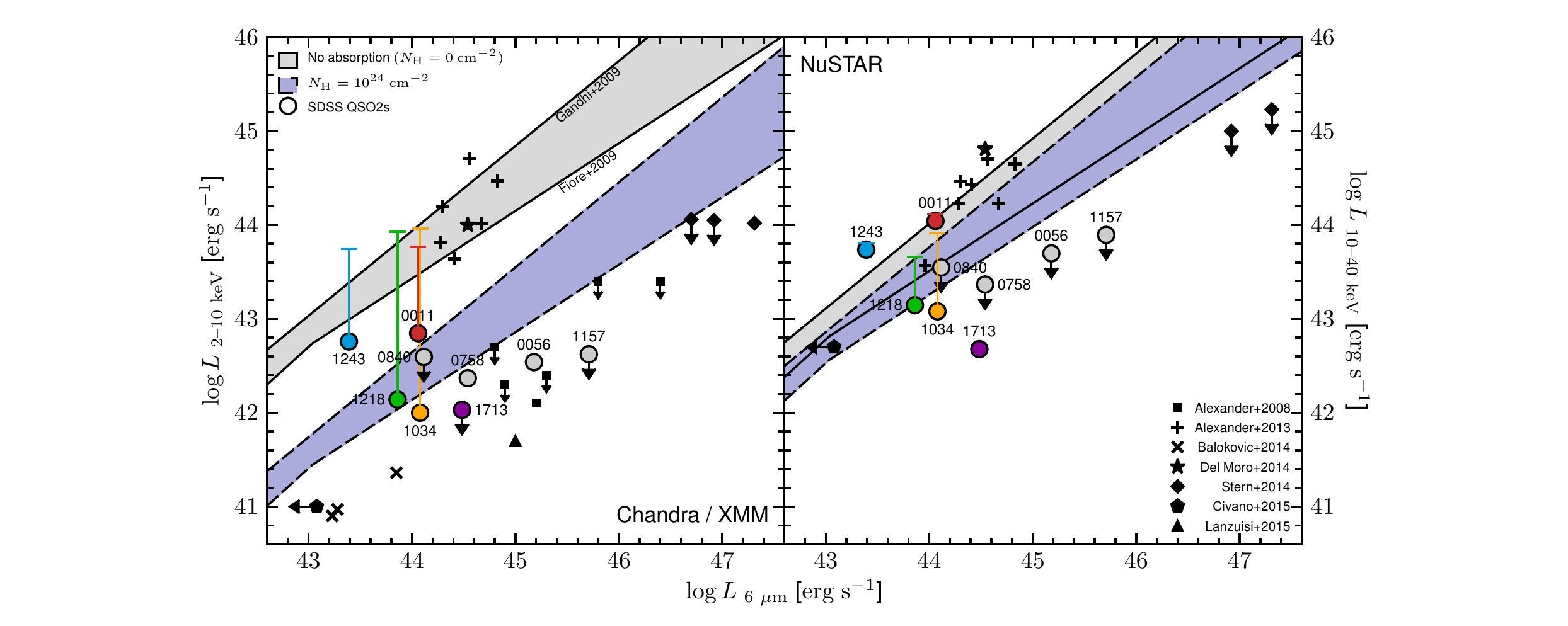}
\caption{Observed (i.e., uncorrected
  for absorption) X-ray luminosity for the rest-frame
  $2$--$10$~keV and $10$--$40$~keV bands (left and right hand panels,
  respectively) versus rest-frame \sixum luminosity (in $\mathrm{\nu}
  L_{\mathrm{\nu}}$ units). The circles indicate the
  \nustar-observed SDSS-selected candidate CTQSO2s presented in this work, L14, and
  G14 ($z=0.05$--$0.49$); colored circles mark the
  \nustar-detected sources. The X-ray luminosities for the
    candidate CTQSO2s are taken
    from best-fitting spectral models where possible. Otherwise, they
    have been determined from photometry, assuming an unabsorbed power law model with
    $\Gamma=0.3$ (as described in Section \ref{nudat}). For the three
    L14 objects (\sdssa, 0056+0032 and 1157+6003), the values
    have been adjusted for consistency with this work.
Other \nustar-observed objects are shown,
  including: \nustar extragalactic survey AGNs
  \citep[`$+$'  symbols; $z=0.02$--$2.92$;][]{Alexander13}, three CT
  Seyfert~2 AGNs \citep[`$\times$' symbols; $z\approx 0.01$;][]{Balokovic14}, a heavily obscured quasar identified in the
  ECDFS field \citep[star; $z\approx2$;][]{DelMoro14}, three luminous and heavily
  obscured {\it WISE}-selected AGNs \citep[diamonds;
  $z\approx2$;][]{Stern14}, and a CT AGN identified in the
  COSMOS field (pentagon; $z=0.044$; F~Civano
  et al., submitted). For the latter object, we show an upper
  limit in \Lsixum, since we have assumed that the mid-IR emission is
  AGN-dominated. Additionally, for the $2$--$10$~keV band we
  compare with sources studied at $<10$~keV with \chandra or \xmm: the candidate CT quasars presented in
  \citeauthor{Alexander08} (\citeyear{Alexander08}; squares;
  $z\approx2$), and a candidate heavily CT AGN identified in
    the COSMOS field \citep[triangle; $z=0.35$;][]{Lanzuisi15b}.
For four of the \nustar-observed candidate CTQSO2s, vertical lines
indicate the intrinsic (i.e., corrected for absorption) X-ray luminosities obtained from X-ray
spectral analyses.
We compare all of the data with two intrinsic relations for the
$2$--$10$~keV band (solid black lines),
those of \citet{Fiore09} and \citet{Gandhi09}. Following L14 and
\citet{Stern14}, the relations have been extrapolated to 
$10$--$40$~keV assuming $\Gamma=1.8$, and the dashed lines show the
effect of absorption by $N_{\rm H}=10^{24}$~\nhunit gas. 
The different X-ray:mid-IR ratios for the
\citet{Fiore09} and \citet{Gandhi09}
relations means that the former provides a more conservative estimate
of the CT absorption threshold.
The majority of the \nustar-observed candidate
  CTQSO2s have low X-ray:mid-IR ratios, suggesting CT levels of
  photoelectric absorption.}
\label{lx_lmir}
\end{figure*}

It is well-established that there is a tight relation
between the mid-IR and intrinsic X-ray luminosities of AGNs \citep[e.g.,][]{Lutz04,Fiore09,Gandhi09,Lanzuisi09,Mateos15,Stern15}. 
Mid-IR emission can therefore provide an indirect estimate of the
intrinsic AGN power, especially useful when heavy absorption in the
X-rays makes this information challenging to obtain
\citep[e.g.,][]{Vignali10,Alexander08,LaMassa09,LaMassa11,Goulding11,Lanzuisi15b}.
Following the approach used for other \nustar studies of faint,
obscured AGNs (L14; \citealt{Stern14}), in Figure \ref{lx_lmir} we
compare the observed X-ray:mid-IR
luminosity ratios with
intrinsic ratios for unobscured AGNs and those corresponding to
X-ray absorption due to dense obscuring material ($N_{\rm
  H}=10^{24}$~\nhunit), for both the low ($2$--$10$~keV) and high
($10$--$40$~keV) energy X-ray regimes. We show the full sample of nine
\nustar-observed SDSS-selected candidate CTQSO2s, including the five
presented in this work, the three from L14 and
the one in G14.
The X-ray luminosities (\lxobs) are observed values (i.e., uncorrected
for absorption), and the \sixum luminosities
(\Lsixum, in $\mathrm{\nu} L_{\mathrm{\nu}}$ units) are intrinsic values (i.e., corrected for dust extinction
occuring in the system) for the AGN determined through SED modeling
(Section \ref{multiwav}), and both correspond to the values provided in Table
\ref{xray_lums}.
We note that for a large fraction of CT AGNs, potentially $\approx50\%$ in the case
of local CT AGNs, we expect significant
absorption in the mid-IR \citep[e.g.,][]{Bauer10,Goulding12}. We have
partially addressed this through dust corrections which are included in the
SED modeling (Section \ref{multiwav}). These corrections are
  small, however, with the luminosities changing by factors ranging
  from $1.03$ to $1.46$ (with a median of $1.17$).
For the four candidate CTQSO2s with constrained intrinsic X-ray luminosities (\lxint), we plot the \lxint values obtained from X-ray
spectral analyses (see L14, G14, and Sections \ref{sdss1218_spectral} and
\ref{sdss1243_spectral} of this work). We conservatively
adopt intrinsic X-ray luminosities from the models with lower best-fit column
densities (e.g., \modelM in the case of \sdssg and the absorbed power law model
in the case of \sdssi).

The two intrinsic relations utilized for comparison are those of \citet{Fiore09} and
\citet{Gandhi09}, which were both computed at $2$--$10$~keV. In the
case of the \citet{Gandhi09} relation, we adjust the
\twelveum (the mid-IR wavelength at which the relation
was computed) $\mathrm{\nu} L_{\mathrm{\nu}}$ luminosities downwards by $7\%$ to obtain \sixum
luminosities, based on the \citet{Assef10} AGN template.
The two relations predict slightly different X-ray:mid-IR ratios
at low luminosities and diverge further towards higher luminosities, which
is partly due to the different luminosity ranges over which the two
relations were calibrated, but also reflects the uncertainty in such
relations. Comparison to both allows us to account for systematic
effects in the derivation of these relations.
We extrapolate the relations to the
$10$--$40$~keV band assuming $\Gamma=1.8$ (typical value for AGNs;
e.g., \citealt{Alexander13}).
An advantage of using $10$--$40$~keV 
X-ray luminosities (\Lhardobs), as opposed to $2$--$10$~keV luminosities
(\Lsoftobs), is that contamination from processes other
than AGN continuum emission is negligible in this high-energy band.
However, the suppression of the X-ray emission by absorbing gas is less
dramatic in the $10$--$40$~keV band,
as demonstrated by the relative normalization of the $N_{\rm
  H}=10^{24}$~\nhunit lines in the left and right hand panels of
Figure \ref{lx_lmir}, which were computed assuming a MYTorus model
with $\Gamma=1.8$ and $\theta_{\rm obs}=70^{\circ}$ (following L14).
Absorption by $N_{\rm H}=10^{24}$~\nhunit gas results in a
suppression of the X-ray emission by factors of $\approx20$ and $\approx2$ in the
$2$--$10$~keV and $10$--$40$~keV bands, respectively.
We note that for the four candidate CTQSO2s with 
  \lxint values constrained using X-ray spectral analyses, the intrinsic
luminosities agree more closely with the \citet{Gandhi09} relation
than with the \citet{Fiore09} relation.

In general, the overall sample of candidate CTQSO2s have extremely low $2$--$10$~keV:mid-IR
ratios, with the observed $2$--$10$~keV luminosities a factor of $\gtrsim20$ lower than the
intrinsic relations, suggesting CT absorption. This was already apparent
from $2$--$10$~keV luminosities published in the literature, but here we have demonstrated the $2$--$10$~keV suppression
using our own soft X-ray analysis. A similar conclusion is reached in
the high-energy $10$--$40$~keV band, where six out of nine of the objects have X-ray
luminosities a factor of $\gtrsim2$ lower than the
intrinsic relations, consistent with CT obscuration.
Our sample of SDSS-selected candidate CTQSO2s lies below
the majority of the AGNs detected in the \nustar extragalactic surveys \citep{Alexander13},
including a heavily obscured quasar detected in ECDFS
(NuSTAR~J033202--2746.8; $z\approx 2$; \citealt{DelMoro14}).

Of the five new objects presented in this work, there is one, \sdssi,
which does not appear compatible with CT absorption based on this
indirect analysis. For this object, the low \nh implied by
the relatively high X-ray:mid-IR ratios is incongruous with the
direct constraints from X-ray spectral modeling (Section \ref{sdss1243_spectral}), which
suggest $N_{\rm H}\gtrsim 10^{24}$~\nhunit. 
A similar case where the \nh values inferred from X-ray spectral
modeling and the X-ray:mid-IR ratio do not agree is that of
NuSTAR~J033202--2746.8 (star symbol in Figure \ref{lx_lmir}; \citealt{DelMoro14}). Despite the large column density
measured for this source ($N_{\rm H}\approx 6\times
10^{23}$~\nhunit; \citealt{DelMoro14}), it lies high with respect to
the relations, which may in part be due
to its significant Compton reflection component. It is possible that
a strong reflection component also contributes to the high
X-ray:mid-IR ratio observed for \sdssi, especially given that a pure
reflection spectrum well describes the data (see Section \ref{sdss1243_spectral}).

Of the \nustar targets detected at high energies ($>10$~keV), \sdssh has the most
extreme $10$--$40$~keV:\mir ratio, with a \Lhardobs value suppressed
by a factor of $\approx 35$ with respect to the intrinsic
relations (on average). The fact that the source lies even lower than the
CTQSO2 \sdssd (G14) may be due to some combination of a
heavily CT absorbing column ($N_{\rm H}\gg 10^{24}$~\nhunit) and a less prominent reflection
component. For the non detections, \sdssj and \sdsse, the \Lhardobs upper
limits suggest that if the X-ray faintness is due to absorption, these
sources are likely CT (for \sdsse this only
  applies for the \citealt{Gandhi09} relation). While heavy absorption seems the most likely
explanation for the X-ray faintness of these non detections, we do not
have broad-band X-ray spectral constraints and therefore cannot rule out the possibility of intrinsic
X-ray weakness
\citep[e.g.,][]{Gallagher01,Wu11,Luo14,Teng14}. However, intrinsic
  X-ray weakness is a phenomenon observed for type 1 sources where
  there is an unobscured view of the central nucleus, 
unlike for our QSO2s.

\section{Discussion}
\label{discussion}

In the following sections,
we discuss the possible implications of the extremely high column
densities and corresponding intrinsic luminosities
measured for the \nustar-detected heavily obscured QSO2s presented in this
paper (\sdssg, 1243--0232 and 1713+5729), L14 (\sdssa), and G14
(\sdssd), in the context of the overall quasar population.

\subsection{Heavy Absorption and Powerful X-ray Luminosities}
\label{discussion_1}

\begin{figure}
\centering
\includegraphics[width=0.47\textwidth]{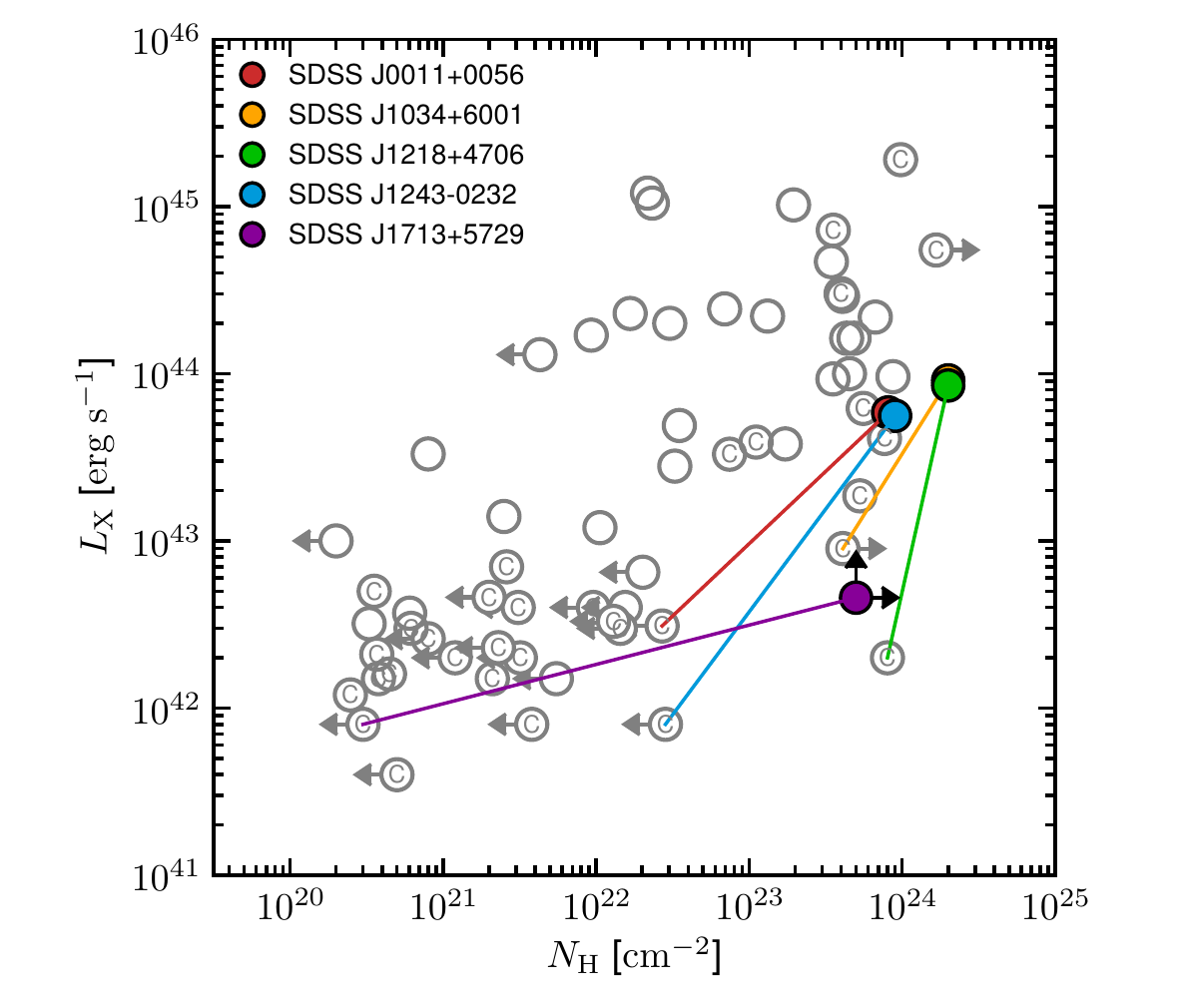}
\caption[]{Intrinsic (i.e.,
  absorption-corrected) rest-frame $2$--$10$~keV luminosity
  (\Lsoftint, or \lx) versus $N_{\mathrm{H}}$ for SDSS-selected QSO2s, as measured from X-ray
  spectral analyses. The grey open circles show the constraints
  in the literature, all directly measured from X-ray spectral
  fitting at soft X-ray energies
  ($<10$~keV; J13; \citealt{LaMassa14}). Sources
  with evidence for being CT, primarily based on the indirect
  X-ray:\oiii luminosity ratio diagnostic, are marked with a ``C''. 
  The colored circles show our constraints for the five
  \nustar-observed candidate CTQSO2s detected at high energies ($>10$~keV), from
  the broad-band \nustar plus soft X-ray spectral analyses presented
  in this study, L14, and G14. The colored lines indicate the significant increase
  in both \lx and \nh for these five objects between the soft X-ray
  constraints in the literature and the \nustar plus soft X-ray studies. We note that for
  \sdssi the increase in \lx shown (blue line) may be an
  overestimate.\footnote{For \sdssi, we measure a significantly higher \chandra flux (by roughly a factor
  of eight) than J13 using the same data. While there is not a
  clear reason for this discrepancy, we note that our measured
  \chandra $3$--$8$~keV flux agrees well with the \nustar flux for the
  same energy band (see Table \ref{xray_lums}).\label{j13gbl1243}}}
\label{lx_nh}
\end{figure}
Figure \ref{lx_nh} shows \nh versus intrinsic (i.e.,
absorption-corrected) X-ray luminosity
for all SDSS-selected QSO2s that have been studied at
low energies ($<10$~keV) with \chandra and \xmm, and have {\it direct}
constraints from X-ray spectral analyses. The intrinsic X-ray luminosities shown
are for the rest-frame $2$--$10$~keV band (\Lsoftint), and are hereafter referred to
as \lx. 
The data are compiled from J13 and
\citet{LaMassa14}. Since these two studies have different
approaches, with the former limiting the spectral analysis to
absorbed power law models and the latter using physically motivated models,
we adopt the \citet{LaMassa14} values where multiple measurements exist.
Overlaid are the five sources which have
$8$--$24$~keV detections with \nustar, for which it is therefore possible to
remeasure \nh and \lx with the addition of the high-energy ($>10$~keV)
data. 
In each case, there is a range of column densities consistent
  with the data. To be conservative, we adopt measured values at the lower end
  of these ranges: e.g., for \sdssg we adopt the \modelM results
  ($N_{\rm H} = 2.0\times 10^{24}$~\nhunit; Section
  \ref{sdss1218_spectral}) and for \sdssi we adopt
  the absorbed power law model results ($N_{\rm H} = 9\times
  10^{23}$~\nhunit; Section \ref{sdss1243_spectral}).
The improvements made with \nustar are illustrated by the colored
lines, which connect the literature constraints prior to \nustar and 
the broad-band, \nustar plus soft X-ray constraints. 

Our \lx and \nh measurements for these five objects are
significantly higher than the constraints in the literature from
spectral modeling of the soft X-ray (\chandra or \xmm) data alone. For the
fainter quasars which have net \chandra ($0.5$--$8$~keV) or \xmm PN
($0.5$--$10$~keV) source counts of $S_{\rm soft}\lesssim15$
(\sdssa, \sdssiShort and \sdsshShort) the soft X-ray constraints
underpredict \nh by factors of $k_{N_{\rm
  H}}\approx30$--$1600$, while for the
brighter sources with $S_{\rm soft}\gtrsim50$ (\sdssd and \sdssgShort) \nh is
underpredicted by factors of $k_{N_{\rm
  H}}\approx2.5$--$5$. 
In general, the intrinsic X-ray luminosities (\lx) measured are $\approx
1$--$2$ orders of magnitude higher with the addition of \nustar data,
which is largely due to the increased absorption correction.
These results have implications for X-ray studies of AGNs at
$z<1$ that lack sensitive high-energy ($>10$~keV) coverage.
For example, on the basis of our results we infer that X-ray data at
$<10$~keV may not reliably identify heavily obscured to CT ($N_{\rm
  H}\gtrsim5\times10^{23}$~\nhunit) AGNs if the photon
counts are low, and the intrinsic luminosities will be 
underestimated. A similar conclusion was reached by \citet{Wilkes13},
who used \chandra and multiwavelength data to investigate the intrinsic
X-ray properties of quasars selected
at low radio frequencies.

The intrinsic X-ray luminosities of our objects (close to $L_{\rm X}=10^{44}$~\ergpersec, which roughly agrees with the $L_{\rm
    X,*}$ value for unobscured AGNs; e.g., \citealt{Hasinger05}) makes them important for population synthesis models of the CXB, since 
  $z\lesssim1.5$ AGNs around this luminosity produce most of the CXB at
  its high energy peak
  \citep[e.g.,][]{Treister05}.\footnote{While the
      \nustar-detected objects all satisfy the classical optical
  quasar luminosity definition (see Sections \ref{definitions} and
  \ref{selection}), based on Figure \ref{lx_nh} they are just below
  the standard `X-ray quasar' luminosity threshold ($L_{\rm
    X}>10^{44}$~\ergpersec), although \sdssd, \sdssgShort and \sdssiShort are
consistent with lying above the threshold for some of the X-ray
spectral model solutions.}  
It is thus useful to consider the \nh distribution and CT
fraction for this class of optically selected QSO2s.

\subsection{The $N_{\rm H}$ Distribution}
\label{discussion_2}

\begin{figure}
\centering
\includegraphics[width=0.47\textwidth]{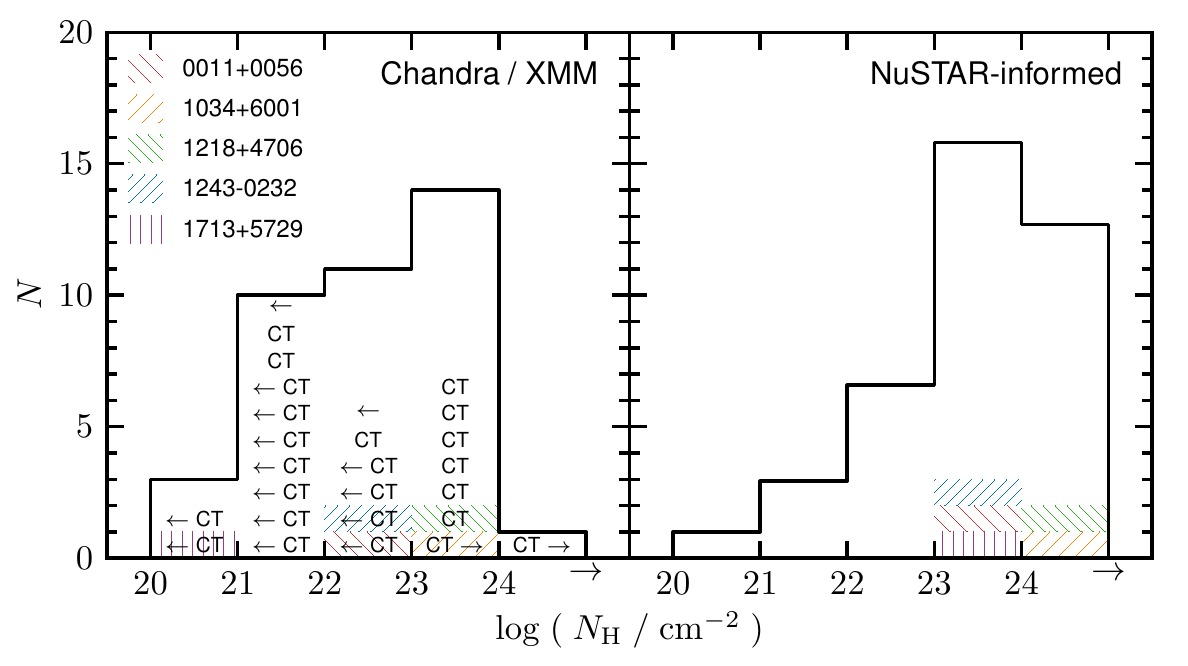}
\caption{The \nh distribution of SDSS-selected QSO2s at $z<0.5$, constructed using
  only direct constraints from X-ray spectral analyses. The five
  \nustar-observed objects with enough $>10$~keV counts for X-ray spectral
  analyses are marked by the hatched regions. Left
  panel: a measurement of the \nh distribution from existing
  soft X-ray ($<10$~keV) \chandra and \xmm constraints. CT candidates,
  identified as such in J13 and \citet{LaMassa14}
  primarily using the indirect X-ray:\oiii luminosity ratio
  diagnostic, are labeled as ``CT''. Sources with
  upper/lower limits in \nh are marked with arrows. Right panel:
  Prediction for the true \nh distribution of SDSS-QSO2s, based on the
  results of this broad-band \nustar plus soft X-ray study; see
  Section \ref{discussion_2}.}
\label{nh_dist}
\end{figure}

In the left panel of Figure \ref{nh_dist} we show the observed \nh distribution for
SDSS-selected QSO2s that
are detected with \chandra and \xmm, and
have {\it direct} constraints at $<10$~keV from X-ray spectral
fitting (J13; \citealt{LaMassa14}). The $39$ objects included have $z<0.5$ and \loiii$>2.5\times10^{8}$~$L_{\rm
  \odot}$, and should therefore be broadly representative of the overall
optically selected QSO2 population (for further details, see Section
\ref{selection}). 
The exclusion of
QSO2s undetected by \chandra and \xmm has a negligible impact since, for the adopted $z$ and \loiii
ranges, there are only three such objects.
On the basis of these data, the column density distribution is relatively flat at $N_{\rm H}=10^{21}$--$10^{24}$~\nhunit,
and there is only one object above $N_{\rm
  H}=10^{24}$~\nhunit. The absorber for this object
  (SDSS~J0939+3553) appears different in nature to those presented in this
  paper, possibly taking the rare form of a geometrically thin toroidal ring \citep{LaMassa14}.

In this work, we have demonstrated that soft X-ray (\chandra and \xmm)
studies can
underpredict the \nh and \lx values of quasars with evidence for CT absorption
based on multiwavelength diagnostics (CT candidates; see Section
\ref{discussion_1} and Figure \ref{lx_nh}). The severity of
the \nh and \lx underpredictions is related to the observed soft X-ray source
photon counts ($S_{\rm soft}$), with the faintly detected sources
suffering larger
underpredictions than the more strongly detected sources. 
To understand the consequences of this for the true \nh
distribution of QSO2s, our result for the \nustar-detected objects can be
extrapolated to the remaining CT
candidates in Figure~\ref{nh_dist}, which were identified as such primarily based on the
X-ray:\oiii luminosity ratio (J13; \citealt{LaMassa14}).
This extrapolation relies on assuming that the \nustar-detected
subsample of five objects are representative of the remaining subsample of $19$ CT
candidates in terms of their absorption properties.
This is a reasonable assumption; the $L^{\rm
    obs}_{\mathrm{X}}/L_{\rm 6\mu m}$ distributions of the two
  subsamples are in agreement (KS test: $p=0.70$), using the X-ray
  luminosities from J13 (except for \sdssi, for which we use our measured
  luminosity; see footnote \ref{j13gbl1243}) and estimating the \sixum luminosities from an
  interpolation between the \wise photometric bands.

To make a prediction for the
true \nh distribution of optically selected QSO2s, we apply an \nh
correction factor ($k_{N_{\rm H}}$) to each of the
$19$ CT candidates in Figure \ref{nh_dist} not observed/detected with
\nustar, informed by our \nustar-measured $k_{N_{\rm H}}$ values (Section \ref{discussion_1}).
For sources with low ($S_{\rm soft}<33$) and high ($S_{\rm soft}>33$) soft X-ray source
counts (using PN counts only in the case of \xmm data)
we draw correction factors at random from flat
distributions between $1.5<\log (k_{N_{\rm H}}) <3.2$ and between
$0.4<\log (k_{N_{\rm H}}) <0.7$, respectively. 
In determining these correction factors we assumed column densities which are
  at the lower end of the range that is consistent with the data (Section \ref{discussion_1}):
  for the three most strongly detected
  sources (\sdssd, 1218+4706, 1243--0232), the lowest best-fit $N_{\rm H}$ values of $(0.9$--$2.0)\times
  10^{24}$~\nhunit are adopted, although the sources are
  consistent with having much larger columns
  ($N_{\rm H} \gtrsim 5\times 10^{24}$~\nhunit); and we assume the \nh
  lower limit for \sdssh ($N_{\rm H}=5\times 10^{23}$~\nhunit). As such, the \nh distribution prediction
  below may provide a lower limit on the CT
  fraction. However, this discussion is ultimately limited by the
  small number of sources detected above $10$~keV with \nustar.

The predicted \nh distribution (averaged over many iterations) is shown in the right
hand panel of Figure \ref{nh_dist}.
This ``\nustar-informed'' $N_{\rm H}$ distribution for optically
selected QSO2s is strongly
skewed towards high columns of $N_{\rm H}>10^{23}$~\nhunit. 
Our predicted
CT fraction ($f_{\rm CT}$), defined here as the ratio of the number of objects with $N_{\rm H}>10^{24}$~\nhunit
to those with $N_{\rm H}>10^{22}$~\nhunit, is
$f_{\rm CT} =36^{+14}_{-12}$~$\%$, where the errors represent binomial
uncertainties only. The full uncertainties are likely to be
larger; considering extreme $k_{N_{\rm H}}$
distributions, where the $k_{N_{\rm H}}$ values assumed are all set equal to either
the highest or lowest values of the ranges measured with \nustar, the uncertainties
on $f_{\rm CT}$ may be larger by a factor of $\approx2$.

The CT fraction
is an important parameter in population
synthesis models of the CXB. In many such models, $f_{\rm CT}$ is treated as a fixed, global
quantity; the \citet[][]{Treister09} model assumes a relatively low value of $15\%$,
while others assume $50\%$ (\citealt{Gilli07}, \citealt{Ueda14}; the quoted
fractions have been adjusted from the original published values to
our adopted definition of $f_{\rm CT}$).
It is possible to estimate $f_{\rm CT}$ using this class of CXB
synthesis model, although 
meaningful constraints are challenging to obtain due to degeneracies
with other parameters (e.g., \citealt{Akylas12}). Fixing the
Compton-reflection strength parameter, \citet{Ueda14} constrain $f_{\rm
  CT}=33$--$62\%$, which
is compatible with our result. 
In other CXB synthesis models, the CT fraction is dependent on physical
  properties of the AGN population; according to the \citet{Draper10} model,
  high CT fractions are associated (beyond the local
  Universe) specifically with black holes accreting at
  a large fraction of their Eddington rate, in broad consistency with
  our findings.

With the \nh distribution in Figure \ref{nh_dist} we have attempted to
provide a prediction using only {\it directly} measured column
densities since analysis of the X-ray spectrum should provide the
``purest'' measurement of the line-of-sight column density, without the
need to make assumptions in comparing emission across very different
wavelength regimes (i.e., using {\it indirect} absorption diagnostics
such as the X-ray:mid-IR, X-ray:\oiii or X-ray:\nev luminosity
ratios).
However, it is worthwhile considering an extreme scenario in which all
of the candidate CTQSO2s in Figure \ref{nh_dist} (labelled ``CT'') are
truly CT; i.e., in which the indirect absorption diagnostics are assumed to be accurate. 
Applying this assumption, the predicted CT fraction is $f_{\rm CT}=65^{+11}_{-13}$~$\%$.
For comparison, \citet{Vignali10} make similar assumptions
using the X-ray:\oiii and X-ray:mid-IR luminosity ratios for a
complete sample
of $25$ SDSS-selected QSO2s at $z\approx 0.5$,
and determine $f_{\rm CT}\approx 50\%$. 
Additionally, \citet{Vignali14} utilize the X-ray:\nev ratio for a sample
of $z\approx0.8$ type 2 AGNs and find $f_{\rm CT}\approx 40\%$.
In the case of Seyfert 2s in the local Universe, \nh distributions have been
constructed for optically selected samples using indirect absorption
diagnostics (primarily the X-ray:\oiii ratio),
predicting a fraction of $f_{\rm CT}\gtrsim 50\%$ for this lower
luminosity AGN population \citep[e.g.,][]{Bassani99,Risaliti99,LaMassa11}.

Indirect absorption diagnostics predict a larger CT
  fraction for $z<0.5$ QSO2s than our \nustar-informed \nh
  distribution. The apparent discrepancy may well be due to indirect
  diagnostics overpredicting the number of CT AGNs. Another
  reconciling factor could be that the quasars unobserved/undetected with \nustar, in general, suffer
even heavier absorption than our detected objects.
Deeper observations at both low (e.g., with {\it Athena};
\citealt{Nandra13}) and high (e.g., with \nustar or {\it Astro-H};
\citealt{Takahashi12}) X-ray energies are needed to reliably distinguish between
the above scenarios, and thus achieve tighter constraints on
$f_{\rm CT}$ for the quasar population.

\section{Summary}
\label{summary}

Sensitive high-energy ($>10$~keV) \nustar observations of five optically selected candidate CTQSO2s have been
presented, along with broad-band X-ray spectral and multiwavelength
analyses. 
Similar studies for a further four such
objects have already been presented in the
literature (L14; G14). The overall sample of nine $z<0.5$ candidate CTQSO2s was selected
primarily on the basis of multiwavelength evidence for absorption by CT ($N_{\rm
  H}>1.5\times 10^{24}$~\nhunit) material along the line-of-sight
(see Section \ref{selection}). Our results are summarized as follows:

\begin{itemize}

\item Of the five recently observed objects, two are
  undetected by \nustar at $8$--$24$~keV (\sdssj and 0840+3838), one is weakly detected 
  (net source counts $S_{\rm 8-24\ keV}= 38.1^{+19.6}_{-18.1}$; \sdssh), and two are strongly detected
  ($S_{\rm 8-24\ keV}\gtrsim 90$; \sdssg and 1243--0232). These represent the first 
  detections of these sources at high X-ray energies; see Section \ref{nudat}.

\item For the two strongly detected targets, spectral modeling of the
  \nustar plus archival soft X-ray data suggests that the
    primary transmitted AGN
    continua are suppressed by extreme levels of photoelectric absorption,
    with $N_{\rm H}\gtrsim 10^{24}$~\nhunit; see Section \ref{xray_spectral}. For the brightest source, \sdssg, the relatively high quality
  spectral analysis using physically motivated models provides strong
  evidence for CT absorption, with a contribution from Compton
  reflection; see Section \ref{sdss1218_spectral}.

\item For \sdssh, the \nustar detection likely represents the first
  identification of directly transmitted emission from the AGN; see
  Section \ref{sdss1713_spectral}. We
  characterize the $3$--$24$~keV spectrum using the \nustar band ratio
  (\brnu) and estimate a high absorbing column density of $N_{\rm
  H}\gtrsim 5\times 10^{23}$~\nhunit; see Section \ref{xray_br}. 
  Notably, the observed $10$--$40$~keV
  luminosity appears to be extremely suppressed, by a factor of
  $\approx35$, with respect to the intrinsic luminosity, suggesting $N_{\rm H}\gg 10^{24}$~\nhunit if
  purely due to absorption; see Section \ref{indirect}.

\item For the non detections, column
  densities of $N_{\rm H}\gtrsim 10^{24}$~\nhunit 
 are inferred by comparing the upper limits in
  observed X-ray luminosity (at rest-frame $2$--$10$~keV and $10$--$40$~keV)  with the intrinsic
  luminosities expected from the mid-IR emission. The majority of
  \nustar-observed candidate CTQSO2s have X-ray:mid-IR ratios
  suggesting CT absorption; see Section \ref{indirect}.

\item For the five objects in the overall \nustar-observed candidate
  CTQSO2 sample that are
  detected at high energies, the column densities and
  intrinsic luminosities measured from spectral analyses are factors
  of $\approx 2.5$--$1600$ and $\approx 10$--$70$ higher, respectively, than soft X-ray
  constraints in the literature; see Section \ref{discussion_1}. 

\item Using direct constraints on absorption for $39$ QSO2s
    studied at X-ray wavelengths, and assuming that
  the \nustar-detected QSO2s are representative of the larger sample with
  evidence for CT absorption, we make a prediction
  for the \nh distribution of optically selected QSO2s. The distribution is
  highly skewed toward large column densities ($N_{\rm
    H}>10^{23}$~\nhunit) and the predicted CT fraction of $f_{\rm CT}
  =36^{+14}_{-12}$~$\%$ is broadly consistent with CXB models. 
  A higher fraction of up to $76\%$ is possible if indirect
  absorption diagnostics are assumed to be accurate; see
  Section \ref{discussion_2}.

\end{itemize}
\section*{Acknowledgements}

We thank the referee for their careful review, which has improved this work. 
We acknowledge financial support from: the Science and Technology
Facilities Council (STFC) grants ST/K501979/1 (G.B.L.), ST/J003697/1 (P.G.), ST/I001573/1 (D.M.A. and A.D.M.); the Leverhulme Trust (D.M.A.); Gemini-CONICYT grant 32120009 (R.J.A.); the ERC Advanced Grant FEEDBACK at the University of Cambridge (J.A.); NSF AST award 1008067 (D.R.B.); the NASA Earth and Space Science Fellowship Program, grant NNX14AQ07H (M.B.); CONICYT-Chile grants Basal-CATA PFB-06/2007 (F.E.B.), FONDECYT 1141218 (F.E.B.), and ``EMBIGGEN'' Anillo ACT1101 (F.E.B.); the Ministry of Economy, Development, and Tourism's Millennium Science Initiative grant IC120009, awarded to The Millennium Institute of Astrophysics, MAS (F.E.B.); Caltech \nustar subcontract
44A-1092750 (W.N.B. and B.L.); NASA ADP grant NNX10AC99G (W.N.B. and
B.L.); the Caltech Kingsley visitor program (A.C.); ASI/INAF grant I/037/12/0–011/13 (A.C., S.P., C.V.); NASA ADAP award NNX12AE38G (R.C.H.); National Science Foundation grant 1211096 (R.C.H.); and Swiss National Science Foundation grant PP00P2\_138979/1 (M.K.).
We thank Andrew Ptak and Jianjun Jia for the useful correspondence.
This work was supported under NASA Contract No.\ NNG08FD60C, and made use of data from the \nustar mission, a project led by the California Institute of Technology, managed by the Jet Propulsion Laboratory, and funded by the National Aeronautics and Space Administration. We thank the \nustar Operations, Software and Calibration teams for support with the execution and analysis of these observations. This research has made use of the \nustar Data Analysis Software (NuSTARDAS) jointly developed by the ASI Science Data Center (ASDC, Italy) and the California Institute of Technology (USA).

Facilities: \chandra, {\it IRAS}, \nustar, SDSS, \spitzer, {\it Swift}, {\it WISE}, \xmm.

\appendix
\label{appendix}

Here we give further information on the individual \nustar-observed candidate CTQSO2s
presented in this paper, namely relevant multiwavelength properties
and features which provide evidence for CT material (Section A.1). In the case of X-ray
properties, this Section focuses on the low energy ($<10$~keV)
\chandra and \xmm data which was
available prior to the \nustar observations. For the broad-band X-ray
constraints incorporating high energy ($>10$~keV) \nustar data, which
generally suggest extreme absorption, see
Section \ref{Results}. 
In addition, we discuss the
identification of a strong \feka line in the \xmm spectrum
of \sdssa, the single \nustar-detection in the exploratory candidate
CTQSO2 sample presented by L14 (Section A.2). Lastly, we
  provide the near-UV to mid-IR photometric data used in the SED
  modeling (Section A.3). \\

\section{A.1. Additional Information for Individual Objects}

\subsection{SDSS J075820.98+392336.0 (z=0.216)}
\label{sdss0758}

Fitting an unabsorbed power law model to the \xmm $0.5$--$10$~keV
data, we measure a flat effective photon index of $\Gamma_{\rm
  eff}=1.1\pm 0.4$, indicative of photoelectric absorption in the
X-ray spectrum. This source is not detected (above
the $2.6$$\sigma$ confidence level) with \nustar at $8$--$24$~keV; see
Section \ref{nudat}.

\subsection{SDSS J084041.08+383819.8 (z=0.313)}
\label{sdss0840}

From {\it HST} WFPC2 (F814W) imaging \citep{VillarMartin12}, the host
has a spiral morphology, with evidence for a tidal feature. 
\citet{Humphrey10} included this object in their integral-field
observations of six SDSS-QSO2s and found spatially extended \oiii and
\oii emission on scales of up to $27$~kpc, consistent with being
powered by AGN activity (e.g., via shocks or
radiation). 
%
%
Using the available \xmm $0.5$--$10$~keV data for this object we
measure $\Gamma_{\rm eff}=0.7\pm0.1$, a low value suggestive of
heavy absorption. This source is a non detection in the \nustar
$8$--$24$~keV data; see Section \ref{nudat}.

\subsection{SDSS J121839.40+470627.7 (z=0.094)}
\label{info_1218}

The $0.5$--$10$~keV \xmm (obsID 0203270201) spectrum is modeled in J13 and 
\citet{LaMassa12a},  
who measure high column densities of $N_{\rm
  H}=8.0^{+5.6}_{-4.1}\times 10^{23}$~\nhunit and
$N_{\mathrm{H}}=(8.7^{+6.7}_{-3.4})\times10^{23}$~\nhunit,
respectively.
J13 measure a strong \feka
feature at $E_{\rm line}=6.4\pm 0.2$~keV with
$\mathrm{EW}_{\mathrm{Fe\ K}\alpha}=1.7^{+2.4}_{-1.4}$~keV, 
consistent with CT absorption.
This target is strongly detected with \nustar at $8$--$24$~keV,
allowing relatively detailed, broad-band spectral modeling which
extends to high energies ($>10$~keV); see
Section \ref{sdss1218_spectral}.

\subsection{SDSS J124337.34--023200.2 (z=0.281)}
\label{info_1243}

Using {\it HST} ACS imaging, \citet{Zakamska06} find that the host galaxy light profile is well fit by a de Vaucouleurs profile,
implying an elliptical morphology. The host
morphology is notably asymmetric. \citet{Zakamska06} find no evidence for extinction in the host
galaxy, suggesting that kpc-scale dust is not obscuring the AGN, and
measure a blue excess in the nucleus which may be due to scattering or starburst emission.

Studying the existing \chandra data, we find an excess of
emission at observed-frame $\approx 5$~keV (i.e., rest-frame $\approx6.4$~keV). When fitting the continuum
emission with a power law and the excess with a Gaussian component, we
measure a rest-frame centroid energy compatible with \feka ($E_{\mathrm{line}}=6.5^{+0.7}_{-0.2}$~keV), and a rest-frame equivalent width of
$\mathrm{EW}_{\mathrm{Fe\ K}\alpha}=2.5^{+4.2}_{-2.4}$~keV. Although the emission
is consistent with $\mathrm{EW}_{\mathrm{Fe\ K}\alpha}\gtrsim1$~keV, which would suggest the presence of CT
material, there are too few photon counts to rule out low equivalent
widths. The object appears to have an
extremely flat spectrum, with $\Gamma_{\rm eff}=-1.1^{+1.2}_{-1.6}$ for the
$0.5$--$8$~keV energy band, indicating strong photoelectric absorption.
This target is strongly detected at $8$--$24$~keV with \nustar, allowing
broad-band X-ray spectral modeling; see
Section \ref{sdss1243_spectral}.

\subsection{SDSS J171350.32+572954.9 (z=0.113)}
\label{sdss1713}


The mid-IR spectrum, as measured with \spitzer-IRS \citep{Sargsyan11},
is AGN-dominated and has evidence for shallow silicate (Si)
absorption at $\approx10$~$\mu$m. The low energy X-ray
  properties of this source are detailed in Section
  \ref{sdss1713_spectral}. To summarise, an extremely steep spectral
  shape at $0.5$--$10$~keV ($\Gamma \approx 3$) suggests that the
  weak \nustar detection at $8$--$24$~keV is the first identification
  of directly transmitted AGN emission from this system.

\section{A.2. An Iron Line in the X-ray Spectrum of SDSS J001111.97+005626.3}
\label{sdss0011}

The $<10$~keV X-ray spectrum of \sdssa was first presented in J13. L14
extended the X-ray analysis to high energies and used
the \nustar/\xmm band ratio to identify heavy, close to CT, absorption
($N_{\rm H}\approx 8\times 10^{23}$~\nhunit). 
L14 did not perform detailed spectral modeling, due to the low source
counts ($\approx 25$ net source counts). However, studying the \xmm $0.5$--$10$~keV
spectrum we find evidence for an excess at observed-frame
$\approx 4.5$~keV (i.e., rest-frame $\approx 6.4$~keV). Modeling the
continuum emission with a power law and the excess with a Gaussian
component, the rest-frame line centroid energy is in good agreement with
that expected for \feka line emission ($E_{\rm line}=6.4\pm
0.1$~keV), and the rest-frame equivalent width is large
($\mathrm{EW}_{\mathrm{Fe\ K}\alpha}=2.9^{+2.5}_{-2.2}$~keV). This
strong \feka emission suggests CT absorption, and it adds confidence to the high column density
measured by L14.

\section{A.3. Near-ultraviolet to Mid-infrared Photometry}

In Table \ref{UVMIR_table} we provide the near-UV to mid-IR
photometric data set for the five \nustar-observed QSO2s presented in
this work, and the one presented in G14 (\sdssd). This 
data set is adopted for the SED modeling in Section \ref{multiwav}.

\begin{table}
\centering
\caption{Near-Ultraviolet to Mid-Infrared Source Properties}
\begin{tabular}{lcccccc} \hline\hline \noalign{\smallskip}
SDSS J & 0758+3923 & 0840+3838 & 1034+6001 & 1218+4706 & 1243-0232 & 1713+5729 \\
\noalign{\smallskip} \hline \noalign{\smallskip}

$u$ (0.355~$\mu$m)$^{a}$ & $18.967\pm 0.025$ & $20.349\pm 0.179$ & $16.139\pm 0.008$ & $18.727\pm 0.030$ & $20.604\pm 0.116$ & $18.721\pm 0.025$ \\
$g$ (0.468~$\mu$m)$^{a}$ & $18.423\pm 0.008$ & $19.166\pm 0.023$ & $14.743\pm 0.002$ & $17.562\pm 0.008$ & $19.334\pm 0.018$ & $17.480\pm 0.006$ \\
$r$ (0.616~$\mu$m)$^{a}$ & $17.792\pm 0.007$ & $18.021\pm 0.014$ & $14.342\pm 0.002$ & $16.843\pm 0.008$ & $18.015\pm 0.010$ & $16.629\pm 0.004$ \\
$i$ (0.748~$\mu$m)$^{a}$ & $17.629\pm 0.008$ & $17.627\pm 0.013$ & $13.871\pm 0.002$ & $16.386\pm 0.008$ & $17.782\pm 0.012$ & $16.133\pm 0.004$ \\
$z$ (0.892~$\mu$m)$^{a}$ & $17.706\pm 0.019$ & $17.171\pm 0.026$ & $13.698\pm 0.004$ & $16.180\pm 0.014$ & $17.391\pm 0.029$ & $16.093\pm 0.009$ \\
{\it WISE} (3.4~$\mu$m)$^{b}$ & $13.847\pm 0.028$ & $14.322\pm 0.029$ & $11.187\pm 0.024$ & $12.592\pm 0.023$ & $14.762\pm 0.040$ & $12.466\pm 0.023$ \\
{\it WISE} (4.6~$\mu$m)$^{b}$ & $12.267\pm 0.024$ & $13.549\pm 0.035$ & $10.016\pm 0.021$ & $11.448\pm 0.021$ & $14.348\pm 0.063$ & $11.060\pm 0.021$ \\
{\it WISE} (12~$\mu$m)$^{b}$ & $8.659\pm 0.022$ & $10.013\pm 0.041$ & $6.295\pm 0.014$ & $8.242\pm 0.019$ & $11.270\pm 0.155$ & $7.242\pm 0.015$ \\
{\it Spitzer} (3.6~$\mu$m)$^{c}$ & $-$ & $-$ & $-$ & $-$ & $279.100\pm 3.333$ & $-$ \\
{\it Spitzer} (4.5~$\mu$m)$^{c}$ & $-$ & $-$ & $-$ & $-$ & $258.600\pm 3.668$ & $-$ \\
{\it Spitzer} (5.8~$\mu$m)$^{c}$ & $-$ & $-$ & $-$ & $-$ & $280.000\pm 10.640$ & $-$ \\
{\it Spitzer} (8.0~$\mu$m)$^{c}$ & $-$ & $-$ & $-$ & $-$ & $535.300\pm 14.130$ & $-$ \\

\noalign{\smallskip} \hline \noalign{\smallskip}
\end{tabular}
\begin{minipage}[l]{0.94\textwidth}
\footnotesize
\textbf{Notes.} \\
$^{a}$ SDSS DR7 model magnitudes in the AB sinh system, corrected for Galactic extinction. \\
$^{b}$ {\it WISE} magnitudes in the Vega system. We use the gmag
magnitude for \sdssh, and profile-fit magnitudes for the remainder. \\
$^{c}$ {\it Spitzer} 3.8\arcsec\ diameter aperture flux densities in 
units of $\mu$Jy.
\end{minipage}
\label{UVMIR_table}
\end{table}


\end{document}